\documentclass[11pt,a4paper]{article}
\usepackage{jheppub}
\usepackage{subfigure}

\title{Time evolution of Einstein-Maxwell-scalar black holes after a thermal quench}

\author[a]{Qian Chen,}

\author[a]{Zhuan Ning,}

\author[a,b]{Yu Tian,}

\author[c]{Xiaoning Wu,}

\author[d]{Cheng-Yong Zhang}

\author[e]{and Hongbao Zhang,}

\affiliation[a]{School of Physical Sciences, University of Chinese Academy of Sciences, Beijing 100049, China}

\affiliation[b]{Institute of Theoretical Physics, Chinese Academy of Sciences, Beijing 100190, China}

\affiliation[c]{Institute of Mathematics, Chinese Academy of Sciences, Beijing 100190, China}

\affiliation[d]{Department of Physics and Siyuan Laboratory, Jinan University, Guangzhou 510632, China}

\affiliation[e]{Department of Physics, Beijing Normal University, Beijing 100875, China}

\emailAdd{chenqian192@mails.ucas.ac.cn}

\emailAdd{ningzhuan17@mails.ucas.ac.cn}

\emailAdd{ytian@ucas.ac.cn}

\emailAdd{wuxn@amss.ac.cn}

\emailAdd{zhangcy@email.jnu.edu.cn}

\emailAdd{hongbaozhang@bnu.edu.cn}

\abstract{We employ the holographic quench technique to drive Einstein-Maxwell-scalar (EMs) black holes out of equilibrium and study the real-time dynamics therein.
		From the fully nonlinear dynamical simulations, a dynamically unstable Reissner-Nordstr$\ddot{\text{o}}$m anti-de Sitter (RN-AdS) black hole can be scalarized spontaneously after an arbitrarily small quench.
		On the other hand, a dynamically stable scalarized black hole can be descalarized after a quench of sufficient strength.
		Interestingly, on the way to descalarization, the scalarized black hole behaves like a holographic superfluid, undergoing a dynamical transition from oscillatory to non-oscillatory decay.
		Such behaviors are related to the spectrums of quasi-normal modes of scalarized black holes, where the dominant mode migrates toward the imaginary axis with increasing quench strength.
		In addition, due to the $\mathbb Z_{2}$-symmetry preserved by the model, the ground state is degenerate.
		We find that there exists a threshold for the quench strength that induces a dynamical transition of the gravitational system from one degenerate ground state to the other.
		Near the threshold, the gravitational system is attracted to an excited state, that is, a RN-AdS black hole with dynamical instability. 
}

\arxivnumber{}

\begin{document}
	
\maketitle

\section{Introduction}\label{sec:I}
As a toy model, the EMs theory has been extensively studied in gravity and holography. 
Different from the well-known holographic superconducting or superfluid models \cite{Hartnoll:2008vx,Hartnoll:2008kx,Herzog:2008he}, where the charged scalar field is minimally coupled with the Maxwell field,
the EMs models contain a neutral scalar field that interacts with the Maxwell field through a non-minimal coupling function \cite{Astefanesei:2019pfq}.
According to the properties of the coupling function, whether it allows the RN black hole of electrovacuum to solve the given model, these EMs models can generally be divided into two categories.

For the case where there is a linear term of the scalar field in the coupling function, which prohibits the existence of RN black holes of electrovacuum, the corresponding models are generally referred to as the EM-dilaton models with the scalar field, ususlly also called the dilaton field, describing the dilaton behavior of extra dimensions along the four-dimensional spacetime \cite{Kaluza:2018,Klein:1926tv,Cremmer:1978ds}.
Although there are no RN black hole solutions, such models allow the existence of new charged black holes with a non-trivial scalar field configuration, which exhibit some characteristics different from RN black holes, such as the charge-to-mass ratio that can exceed unity \cite{Gibbons:1987ps,Garfinkle:1990qj,Delgado:2016zxv}.
At the linear level, these black holes endowed with scalar hair are verified to be dynamically stable for generic values of the dilaton coupling \cite{Ferrari:2000ep,Zhang:2015jda,Brito:2018hjh,Blazquez-Salcedo:2019nwd,Astefanesei:2019qsg}.
At the nonlinear level, the real-time dynamics of individual black holes under disturbance were numerically studied \cite{Zhang:2021edm,Zhang:2021ybj}, where the perturbed scalar field exhibits oscillation behavior with damping amplitude.
On the other hand, the fully nonlinear dynamical simulations of binary black hole mergers were implemented \cite{Hirschmann:2017psw}, revealing that the differences with respect to waveforms in General Relativity and EM-dilaton are only significant for large charges.
To study the dynamics of binary black holes with arbitrary mass ratios, the post-Newtonian approximation was also used \cite{Julie:2017rpw,Khalil:2018aaj}.
In asymptotically AdS spacetime, due to the rich phase structure, the EM-dilaton models have been widely used in holography to probe the physical properties of QCD \cite{DeWolfe:2010he,DeWolfe:2011ts,Cai:2012xh,He:2013qq,Rougemont:2015ona,Knaute:2017lll,Giataganas:2017koz,Cai:2022omk}.

On the other hand, for the models to allow the existence of RN black holes of electrovacuum, the coupling function must be dominated by the higher order terms of the scalar field.
In this case, in addition to the RN black hole solutions with a trivial scalar field, there still exists a series of charged black holes with scalar hair.
Furthermore, these states in equilibrium exhibit distinct dynamical properties depending on the specific form of the coupling funtion.
For the case where the coupling function is dominated by a quadratic term, the RN black hole with a near-extremal configuration is dynamically unstable \cite{Myung:2018vug,Myung:2018jvi,Guo:2021zed} and undergoes a dynamical transition under arbitrarily small perturbations, spontaneously evolving into a thermodynamically favored scalarized black hole \cite{Herdeiro:2018wub,Fernandes:2019rez,Fernandes:2019kmh,Zhang:2021etr,Luo:2022roz}.
Such a similar spontaneous scalarization phenomenon can also be triggered by the geometric invariant source such as the Gauss-Bonnet invariant \cite{Doneva:2017bvd,Silva:2017uqg,Antoniou:2017acq,Cunha:2019dwb,Dima:2020yac,Herdeiro:2020wei,Berti:2020kgk}.
More generally, the spontaneous scalarization of compact objects in gravity has also been revealed \cite{Damour:1993hw,Brihaye:2019puo,Peng:2019qrl}. 

In stark contrast to the above, the RN black holes are always stable at the linear level in the models with quartic coupling.
Interestingly, in this case, there are two branches of scalarized black holes with diametrically opposite dynamical properties, one of which is linearly stable and the other possesses a single unstable mode \cite{Blazquez-Salcedo:2020nhs,LuisBlazquez-Salcedo:2020rqp}.
For such a gravitational system, the two linearly stable black holes are in two local ground states and the other dynamically unstable black hole is in an excited state.
Through the fully nonlinear accretion mechanism of the scalar field, the two local ground states can be dynamically interconverted by crossing the excited state acting as a dynamical barrier \cite{Zhang:2021nnn,Zhang:2022cmu,Jiang:2023yyn,Chen:2023eru}.
To trigger the dynamical transition, the accretion strength needs to exceed a threshold, near which a class of critical phenomena occurs, analogous to the critical gravitational collapse \cite{Choptuik:1992jv,Choptuik:1996yg,Brady:1997fj,Bizon:1998kq,Gundlach:2007gc}.
Similar critical scalarization and descalarization phenomena also occur in the case of non-minimal high-order coupling of the scalar field to gravity \cite{Liu:2022fxy}.

For the models of quadratic coupling, the triggering mechanism of the spontaneous scalarization in the current research work is limited to the spatial perturbations of the scalar field configuration, which naturally excites the unstable modes of the RN black holes with dynamical instability.
In asymptotically AdS spacetime, there exists a class of holographic techniques that drives the system out of equilibrium, the quantum or thermal quench, which can be regarded as a disturbance based on time domain.
Such a quench mechanism has been widely used to study the dynamical behaviors of holographic systems away from equilibrium, such as thermalization \cite{Janik:2006ft,Chesler:2008hg,Bhattacharyya:2009uu,Chesler:2009cy,Buchel:2014gta}, symmetry breaking \cite{Bhaseen:2012gg,Basu:2012gg,Gao:2012aw,Garcia-Garcia:2013rha,Bai:2014tla,Rangamani:2015agy,Chen:2022vag,Chen:2022tfy} and periodic driving \cite{Li:2013fhw,Yang:2023dvk}.
In this paper, through the fully nonlinear dynamical simulations, we study the real-time dynamics of spontaneous scalarization induced by a thermal quench, revealing the relationship between the scalarization rate and the quench strength for the RN-AdS black holes with a single unstable mode.
On the other hand, we investigate the response of the scalarized black holes with dynamical stability to the thermal quench.
Since the quench opens the closed AdS gravitational system and injects energy into it, the descalarization phenomenon occurs after a quench of sufficient strength.
Such a transition is analogous to the holographic superfluid phase transition from a superfluid state to a normal state under a thermal quench \cite{Bhaseen:2012gg}.
In addition, the dynamical phase diagram of the scalarized black hole is also revealed, which contains three regions that completely resemble a holographic superfluid, implying the potential connection between the EMs scalarized black hole and the holographic superfluid.

The critical dynamical phenomena discovered so far only exist in the gravity model with two local ground states \cite{Zhang:2021nnn,Zhang:2022cmu,Jiang:2023yyn,Chen:2023eru,Liu:2022fxy}.
For the EMs model with quadratic coupling here, the scalarized black hole is the only global ground state.
However, it is degenerate in the case where the $\mathbb Z_{2}$-symmetry is preserved.
We find that the holographic quench is an efficient mechanism for triggering the dynamical transition of the gravitational system between the two degenerate ground states.
After a quench around a specific strength, the gravitational system will stay on a critical RN-AdS black hole with dynamical instability as an excited state for a period of time in the dynamical intermediate process, and then continue to evolve towards a scalarized black hole with a positive or negative scalar condensation, depending on whether the actual quench strength is greater than or less than the critical strength.

The organization of the paper is as follows.
In section \ref{sec:E}, we give a brief description of the EMs models with quadratic coupling, where the field equations, Ward-Takahashi identity and phase diagrams are revealed.
In section \ref{sec:Q}, we impose a fast thermal quench on the gravitational system by specifying a time dependency on the external source of the scalar field.
Under such a quench, the phenomena of spontaneous scalarization of dynamically unstable RN-AdS black holes and continuous descalarization of dynamically stable scalarized black holes are revealed, respectively.
In addition, the dynamical phase diagram during the descalarization process is also investigated.
In section \ref{sec:Cb}, through the dichotomy, we keep approaching the critical quench strengths to reveal the critical behaviors of the gravitational system during the scalarization and descalarization.
Finally, in section \ref{sec:C}, we conclude the paper by a summary and an outlook.

\section{EMs theory}\label{sec:E}
In this section, we introduce the EMs theory from the setup of the gravity model, the Ward-Takahashi identity and the phase diagrams respectively.

\subsection{Gravity model}
The general 4-dimensional EMs gravity with a negative cosmological constant is described by the following action
\begin{equation}
	S=\frac{1}{2\kappa^{2}_{4}}\int d^{4}x\sqrt{-g}\left[R-2\Lambda-\frac{1}{4}f(\phi)F_{\mu\nu}F^{\mu\nu}-\nabla_{\mu}\phi\nabla^{\mu}\phi-m^{2}\phi^{2}\right],\label{eq:action}
\end{equation}
where $R, F_{\mu\nu}, \phi$ represent the Ricci scalar curvature, Maxwell field strength tensor and a real scalar field, respectively.
To work in units of AdS radius, the cosmological constant is set to be $\Lambda=-3$ for convenience.
We consider a massive scalar field with the mass squared $m^{2}=-2$ to respect the Breitenlohner-Freedman (BF) bound \cite{Breitenlohner:1982jf}.
The non-minimal coupling function, which describes the interaction between the real scalar field and the Maxwell field, is assumed to be an exponential dependence dominated by a quadratic term
\begin{equation}
	f(\phi)=e^{\alpha\phi^{2}},\label{eq:coupling}
\end{equation}
where $\alpha$ is a positive coupling constant.
Some other forms of the coupling function, such as $1+\alpha\phi^{2}$, have also been studied, which qualitatively leads to the same conclusions.

From the variation of the action with respect to the metric tensor, one can obtain the Einstein equation
\begin{equation}
	R_{\mu\nu}-\frac{1}{2}Rg_{\mu\nu}=-\Lambda g_{\mu\nu}+T^{M}_{\mu\nu}+T^{\phi}_{\mu\nu}
\end{equation}
with the stree-energy tensors of the Maxwell and scalar fields
\begin{subequations}
	\begin{align}
		T^{M}_{\mu\nu}&=\left(\frac{1}{2}g^{\rho\sigma}F_{\mu\rho}F_{\nu\sigma}-\frac{1}{8}F_{\alpha\beta}F^{\alpha\beta}g_{\mu\nu}\right)f(\phi),\\
		T^{\phi}_{\mu\nu}&=\nabla_{\mu}\phi\nabla_{\nu}\phi-\frac{1}{2}\left(\nabla_{\alpha}\phi \nabla^{\alpha}\phi+m^{2}\phi^{2}\right)g_{\mu\nu}.\label{notation}
	\end{align}
\end{subequations}
The equations of motion for the scalar and Maxwell fields can also be derived through the variation of the action with respect to the corresponding matter fields, as follows
\begin{subequations}
	\begin{align}
		\nabla^{\mu}\nabla_{\mu}\phi&=\frac{1}{8}\frac{df(\phi)}{d\phi}F_{\mu\nu}F^{\mu\nu}+m^{2}\phi,\label{eq:S_E}\\
		\nabla_{\nu}\left[f(\phi)F^{\nu\mu}\right]&=0.\label{eq:M_E}
	\end{align}
\end{subequations}

Obviously, for the quadratic coupling (\ref{eq:coupling}), the RN-AdS black holes with the vanishing scalar field solve the above field equations.
From the equation (\ref{eq:S_E}), one can observe that the coupling constant $\alpha$ combined with the Maxwell invariant contributes to the effective mass of the scalar field.
Although the mass squared $m^{2}$ already satisfies the BF bound in AdS$_{4}$ to ensure the stability of the scalar field on the AdS boundary, some specific structures in the interior of the spacetime may still impose some additional constraints on the effective mass of the scalar field.
For an extremal RN-AdS black hole, such an effective mass squared at the event horizon is a function of the coupling constant $\alpha$,  the radius of the event horizon $r_{h}$ and the mass squared of the scalar field $m^{2}$, namely $u^{2}_{\text{eff}}(\alpha,r_{h},m^{2})$.
Since the near-horizon geometry of an extremal RN-AdS black hole with spherical topology is the direct product of AdS$_{2}$ with a sphere $S^{2}$, the near-horizon instability of the scalar field is induced when the effective mass squared at the event horizon $u^{2}_{\text{eff}}(\alpha,r_{h},m^{2})$ violates the BF bound in AdS$_{2}$, leading to the emergence of spontaneous scalarization.

In order to simulate the fully nonlinear dynamical process of a black hole, which requires us to numerically solve the above time-dependent field equations, the ingoing Eddington-Finkelstein metric ansatz compatible with spherical symmetry is adopted \cite{Chesler:2013lia}
\begin{equation}
	ds^{2}=-2W(t,r)dt^{2}+2dtdr+\Sigma(t,r)^{2}d\Omega^{2}_{2},\label{eq:ansatz}
\end{equation}
where $d\Omega^{2}_{2}$ represents the line element of a unit radius $S^{2}$.
For the scalar and Maxwell fields, we require $\phi=\phi(t,r)$ and take the gauge $A_{\mu}dx^{\mu}=A(t,r)dt$.
The corresponding integral constants required to solve the differential equations can be extracted from the asymptotic behaviors of the field variables near the AdS boundary
\begin{subequations}
	\begin{align}
		\phi&=\phi_{1}r^{-1}+\phi_{2}r^{-2}+o(r^{-3}),\\
		A&=\mu-Qr^{-1}+o(r^{-2}),\\
		\Sigma&=r+\lambda-\frac{1}{4}\phi^{2}_{1}r^{-1}+o(r^{-2}),\\
		W&=\frac{1}{2}(r+\lambda)^{2}+\frac{1}{2}-\frac{1}{4}\phi^{2}_{1}-d_{t}\lambda-Mr^{-1}+o(r^{-2}).
	\end{align}\label{eq:BC}
\end{subequations}
According to the holographic dictionary \cite{Klebanov:1999tb}, the free parameter $\phi_{1}$ is the source of the scalar field, which is used to quench the gravitational system in our work, and the  response $\phi_{2}$ cannot be determined by the near-boundary analysis, whose value depends on the bulk solution.
The pure gauge $\mu$ is set to zero without loss of generality.
The symbols $Q$ and $M$ represent the electric charge and Arnowitt-Deser-Misner (ADM) mass \cite{Abbott:1981ff}, respectively.
The reparameterization freedom $\bar{r}\rightarrow r+\lambda$, under which the form of the metric is invariant, allows us to fix the apparent horizon at a time-independent radial position during a dynamical process.
By the area of the apparent horizon, the entropy density of a black hole is defined as 
\begin{equation}
	s=2\pi\Sigma^{2}(r_{h}),
\end{equation}
where $r_{h}$ stands for the radius of apparent horizon.
{For a black hole in equilibrium, the temperature can be easily obtained from the surface gravity or Euclidean formalism 
\begin{equation}
	T=\frac{1}{2\pi}d_{r}W(r_{h}).
\end{equation}}

\subsection{Ward-Takahashi identity}
From holography, such a gravity model (\ref{eq:action}) is dual to a relevant deformation of strongly coupled conformal field theory with a scalar operator and a conserved current, residing on the AdS boundary.
In order to effectively describe the boundary theory from the gravity theory, some boundary terms need to be introduced to renormalize the bulk action \cite{Gibbons:1976ue,Bianchi:2001kw,Elvang:2016tzz}
\begin{equation}
	2\kappa^{2}_{4}S_{\text{reg}}=\int_{M} dx^{4}\sqrt{-g}\mathcal{L}+2\int_{\partial M}dx^{3}\sqrt{-\gamma}K-\int_{\partial M}d^{3}x\sqrt{-\gamma}\left(4+{R}[\gamma]+\phi^{2}\right),
\end{equation}
where $\mathcal{L}$ is the bulk Lagrangian density, ${R}[\gamma]$ is the Ricci scalar associated with the boundary induced metric $\gamma_{\mu\nu}$ and $K$ is the trace of extrinsic curvature $K_{\mu\nu}=\gamma^{\sigma}_{\mu}\nabla_{\sigma}n_{\nu}$ with $n_{\nu}$ the outward normal vector field to the boundary.
Under the premise that the bulk field equations are satisfied, the variation of the renormalized on-shell action has the form
\begin{equation}
	\kappa^{2}_{4}\delta S_{\text{ren}}=\int_{\partial M}d^{3}x\sqrt{-\gamma_{(0)}} \left(-\frac{1}{2}\left\langle T_{ij}\right\rangle\delta\gamma_{(0)}^{ij}+\left\langle J^{i}\right\rangle\delta A_{i(0)}+\left\langle O\right\rangle\delta\phi_{(0)}\right), \\
\end{equation}
where the subscripts ``$(0)$'' denote the external source of the corresponding field variables on the AdS boundary.
{Note that we use Latin letters for boundary coordinates and $\nabla_{i}$ for boundary covariant derivative operators.}
The expectation values $\{\left\langle T_{ij}\right\rangle,\left\langle J^{i}\right\rangle,\left\langle O\right\rangle\}$ of the field operators of the boundary theory are functions of the asymptotic behaviors (\ref{eq:BC}) of the bulk fields on the AdS boundary, where the boundary energy-momentum tensor is defined as 
\begin{equation}
	\begin{aligned}
		\left\langle T_{ij}\right\rangle&=\lim_{r\rightarrow \infty}r\left[R[\gamma]_{ij}-\frac{1}{2}R[\gamma]\gamma_{ij}-K_{ij}-\left(2-K+\frac{1}{2}\phi^{2}\right)\gamma_{ij}\right],\\
		&=\begin{pmatrix}
			2M-\phi_{1}\left\langle O\right\rangle&0&0\\
			0&M&0\\
			0&0&M\text{sin}^{2}\theta
		\end{pmatrix},
	\end{aligned}
\end{equation}
the conserved current and scalar operator yield the following shapes
\begin{subequations}
	\begin{align}
		\left\langle J^{i}\right\rangle&=-\frac{1}{2}\lim_{r\rightarrow \infty}r^{3}n_{\mu}F^{\mu i}=\left(\frac{1}{2}Q,0,0\right),\\
		\left\langle O\right\rangle &=-\lim_{r\rightarrow \infty}r^{2}\left(\phi+ n^{\mu}\nabla_{\mu}\phi\right)=\phi_{2}+\lambda\phi_{1}-d_{t}\phi_{1},
	\end{align}
\end{subequations}
respectively.

According to the AdS/CFT correspondence, every gauge symmetry of the bulk theory induces a global symmetry of the boundary theory, resulting in a constraint on the boundary operators.
For example, in the present setting the $U(1)$ gauge invariance
\begin{equation}
	\delta A_{i}=\nabla_{i}\Theta,\quad \delta \phi=0,
\end{equation}
gives rise to the source-free conserved current equation
\begin{equation}
	\nabla_{i}\left\langle J^{i}\right\rangle=0,\label{eq:conservation1}
\end{equation}
indicating that the electric charge $Q$ is conserved in the dynamical process.
{Without loss of generality, the electric charge $Q$ is fixed to be 1 in what follows unless otherwise stated.}
This is in marked contrast to the case of a charged scalar field with a time-dependent scalar source, which provides a source for the above equation to induce a change in the electric charge over time \cite{Chen:2022vag}.
In order for the electric charge to be conserved under the quench of the charged scalar field, the time-dependent chemical potential needs to be introduced \cite{Bhaseen:2012gg}.
On the other hand, the residual diffeomorphism invariance of the form (\ref{eq:ansatz}) of the metric
\begin{equation}
	\delta \gamma^{ij}=\pounds_{\xi}\gamma^{ij},\quad \delta{A}_{i}=\pounds_{\xi}A_{i},\quad \delta\phi=\pounds_{\xi}\phi,
\end{equation}
where $\pounds_{\xi}$ is the Lie derivative with respect to an arbitrary vector field $\xi^{i}$ tangent to the boundary, gives the Ward-Takahashi identity 
\begin{equation}
	\nabla^{j}\left\langle T_{ij}\right\rangle=F_{ij}\left\langle J^{j}\right\rangle+\left\langle O\right\rangle \nabla_{i}\phi_{1}.\label{eq:conservation2}
\end{equation} 
The response of the energy density $\epsilon=\left\langle T_{tt}\right\rangle$ of the boundary system to the scalar source can be extracted from the only non-trivial $t$-component of the above vector identity
\begin{equation}
	d_{t}\epsilon=-\left\langle O\right\rangle d_{t}\phi_{1},\label{eq:2.18}
\end{equation}
from which one can observe that the energy of the AdS system is conserved during the dynamical process in the absence of a time-dependent external source.

\subsection{Phase diagrams}
In order to reveal the properties of the black hole solutions in equilibrium in this model with the coupling (\ref{eq:coupling}), we first eliminate the time dependence in the field equations to convert the problem to finding the solutions to a system of ordinary differential equations, which is then numerically solved using the Newton-Raphson iteration algorithm with the pseudo-spectral discretization under the source-free boundary condition $\phi_{1}=0$.

\begin{figure}
	\begin{center}
		\subfigure[]{\includegraphics[width=.49\linewidth]{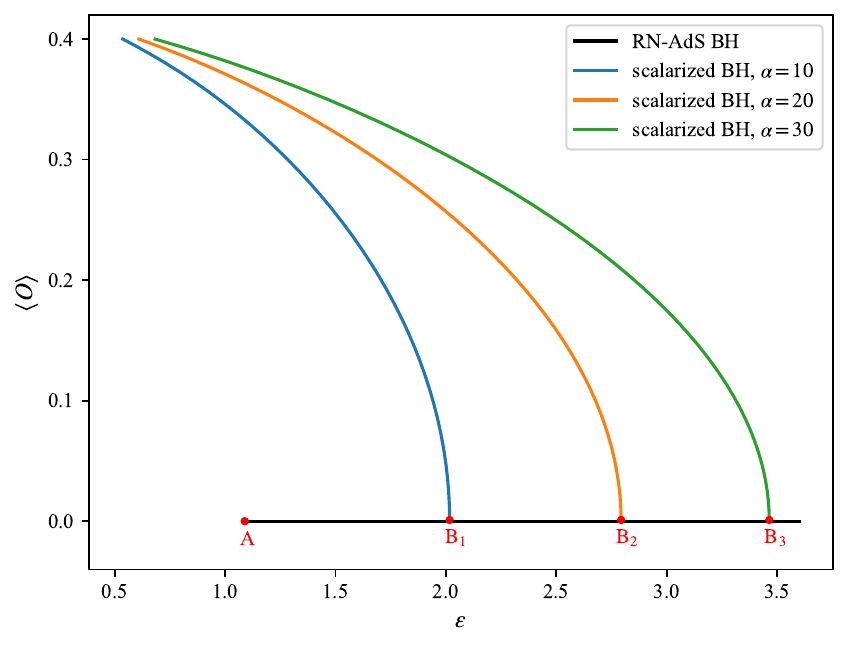}\label{fig:1}}
		\subfigure[]{\includegraphics[width=.49\linewidth]{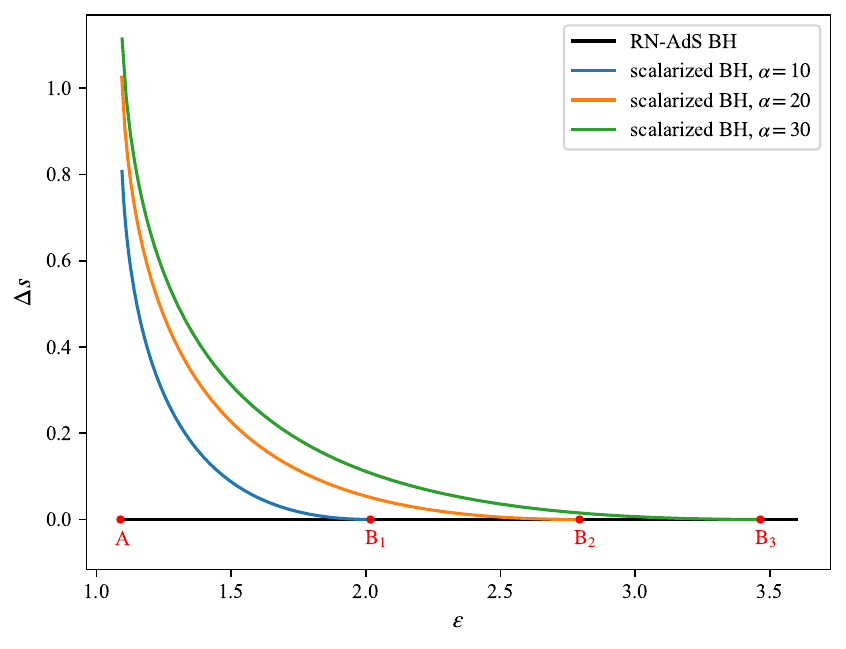}\label{fig:2}}
		\caption{The expectation value of the scalar operator (a) and the difference in entropy density between RN-AdS black holes and scalarized black holes (b) as a function of energy density.
			The black curve represents the branch of RN-AdS black holes.
			Other curves of different colors represent branches of scalarized black holes with different values of the coupling constant $\alpha$.}
		\label{fig:1-2}
	\end{center}
\end{figure}

The resulting phase diagrams are shown in figure \ref{fig:1-2} as the energy density dependence of physical quantities.
As we can see from figure \ref{fig:1}, in addition to the RN-AdS black holes of electrovacuum, for a sufficiently large coupling constant $\alpha$, there exists a branch of scalarized black holes with the non-trivial scalar condensation in the domain of existence of solutions, which intersects with the branch of RN-AdS black holes at point $B$, depending on the coupling constant $\alpha$, and extends to the over-extremal region.
Such a branch of scalarized black holes divides the branch of RN-AdS black holes into two regions, the near-extremal region $AB$ and the far-extremal region, which exhibit distinct dynamical characteristics.
The RN-AdS black holes in the near-extremal region are dynamically unstable as excited states, and can spontaneously evolve into a thermodynamically favored scalarized black hole, as shown in figure \ref{fig:2}, under arbitrarily small perturbations involving any one of the unstable modes.
However, for a gravitational system whose energy density is in the far-extremal region, the RN-AdS black hole with dynamical stability is the global ground state and the only solution.
Thus, if there exists a mechanism that can inject enough energy into a scalarized black hole, then the descalarization phenomenon will occur when the energy density of the gravitational system exceeds that represented by point $B$.
In what follows, we demonstrate that the holographic quench mechanism is up to the task through fully nonlinear dynamics.

\section{Quench dynamics}\label{sec:Q}
In this section, in order to numerically simulate the spontaneous scalarization and continuous descalarization processes induced by a thermal quench, we assign a time dependency to the external source of the scalar field on the static background of a RN-AdS black hole and a scalarized black hole, respectively.
The energy density of the initial gravitational system is set to $\epsilon=1.6$, at this time the RN-AdS black hole is in the near-extremal region and possesses only a single unstable mode.
The coupling constant is fixed to $\alpha=20$.
Requiring that the final state after a quench is in the phase diagram of the previous section, a Gaussian-type quench with mixed frequencies is employed without loss of generality
\begin{equation}
	\phi_{1}=pe^{-\frac{1}{\upsilon}(t-t_{c})^{2}},\label{eq:3.1}
\end{equation}
which rapidly decays to zero at later times.
The symbols $t_{c}$ and $\upsilon$ represent the moment of the peak and the time scale, and are set to $t_{c}=5$ and $\upsilon=0.5$, respectively.
The amplitude $p$ characterizes the quench strength.

\subsection{Spontaneous scalarization}\label{sec:ss}

\begin{figure}
	\begin{center}
		\subfigure[]{\includegraphics[width=.49\linewidth]{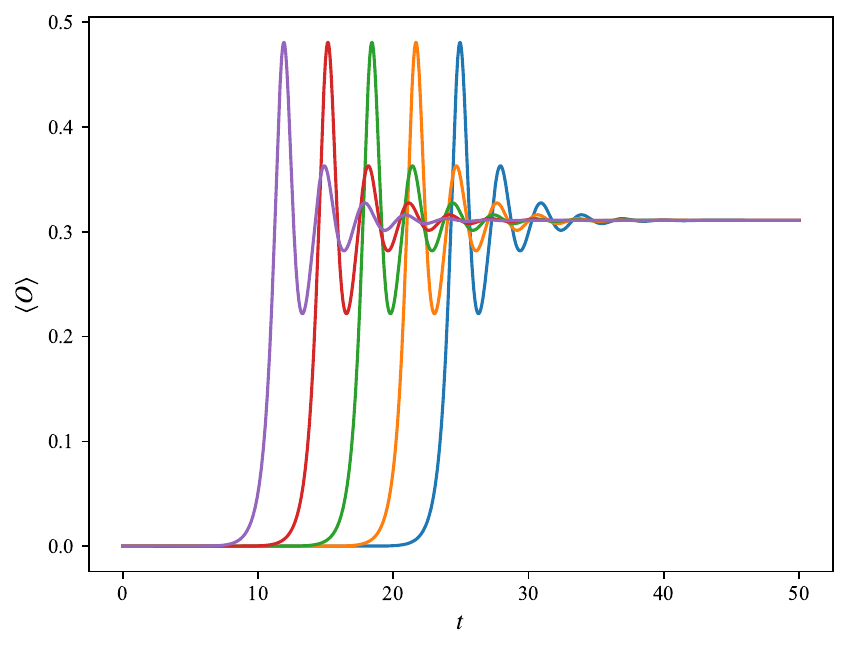}\label{fig:3}}
		\subfigure[]{\includegraphics[width=.49\linewidth]{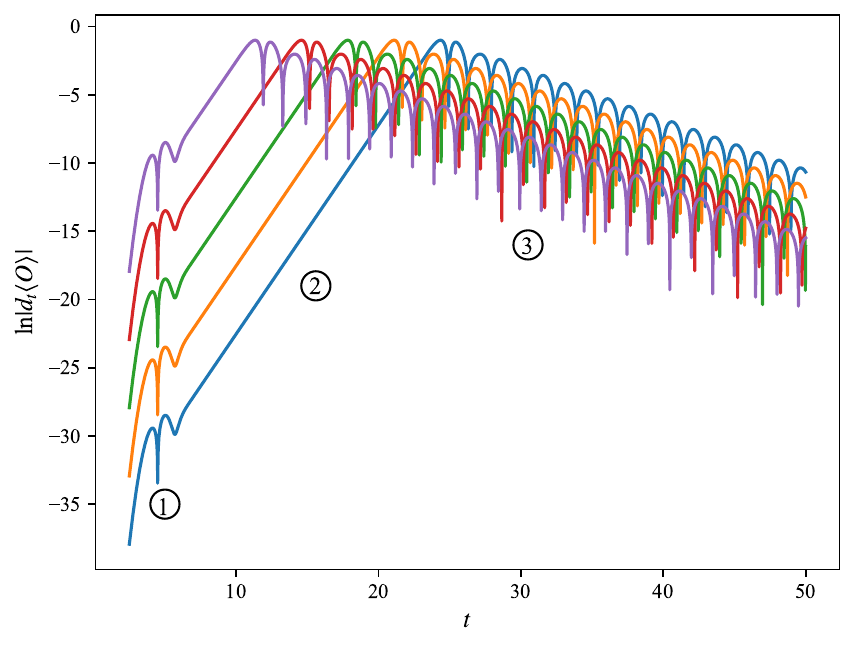}\label{fig:4}}
		\subfigure[]{\includegraphics[width=.49\linewidth]{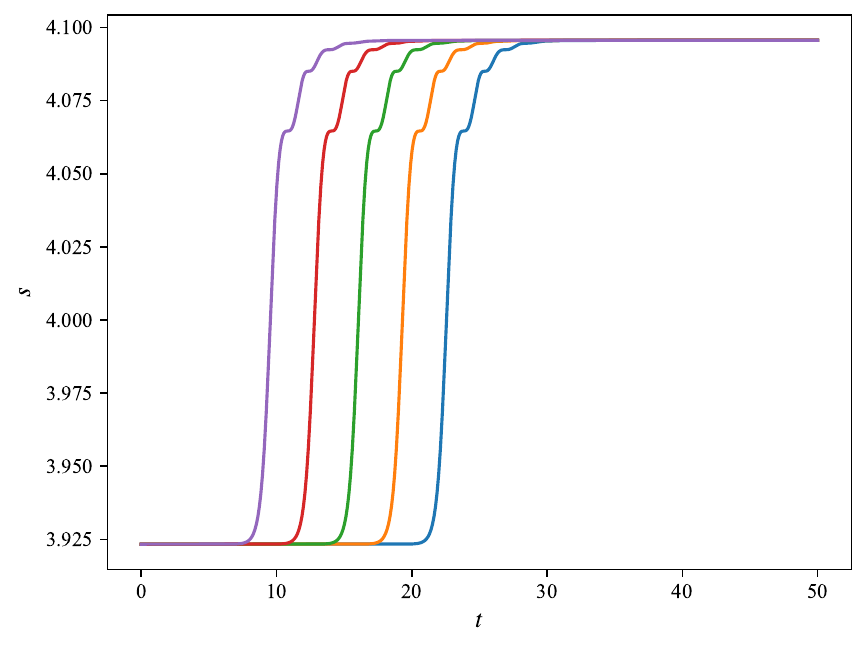}\label{fig:5}}
		\subfigure[]{\includegraphics[width=.49\linewidth]{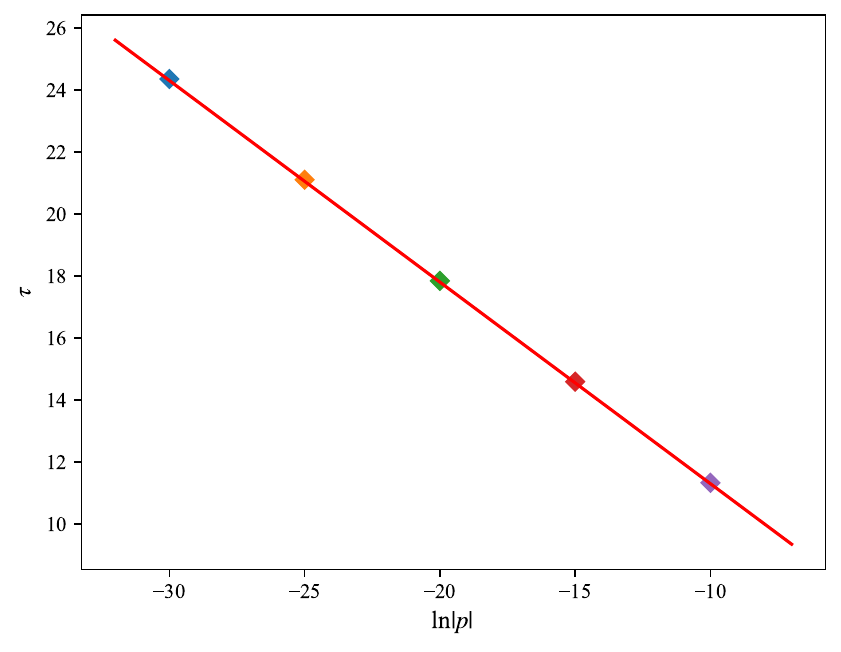}\label{fig:6}}
		\caption{The expectation value of the scalar operator (a), the value of ln$|d_{t}\left\langle O\right\rangle|$ (b) and the entropy density (c) as a function of time in the case where the quench strength $p$ is close to zero. 
			(d):  The time $\tau$ required for the quenched system to converge to the final scalarized black hole with respect to ln$|p|$.
			All the curves and points of the same color in the figure correspond to each other.}\label{fig:3-6}
	\end{center}
\end{figure}

Due to the dynamical instability of the RN-AdS black hole, the spontaneous scalarization can be triggered by arbitrarily small perturbations. 
The resulting real-time dynamics are shown in figure \ref{fig:3-6}, where the quench strengths are set to $p=\{e^{-10},e^{-15},e^{-20},e^{-25},e^{-30}\}$, respectively.
As shown in figure \ref{fig:3}, the quench induces a drastic change in the gravity configuration accompanied by spontaneous generation of the scalar condensation.
After exponential growth, the scalar condensation with oscillatory behavior converges exponentially to a constant, indicating the formation of a scalarized black hole.
By analyzing the dominant mode in the dynamical process, the entire evolution process can be divided into three stages, as shown in figure \ref{fig:4}.
Among them, the first stage is the quench process with a time-dependent scalar source, which slightly changes the energy of the gravitational system and imposes a perturbation on its configuration.
After that, the perturbation excites the single unstable mode of the initial RN-AdS black hole, which grows exponentially, pushing the gravitational system away from the RN-AdS black hole in the second stage.
The slope of the parallel lines in the figure is equal to the imaginary part of the unstable mode.
Finally, in the third stage, the evolving system is captured by a stable scalarized black hole into its linear region, where the non-vanishing real part of the dominant decay mode leads to the oscillatory phenomenon of the scalar condensation at the end of evolution.
In such a scalarization process, the area of the apparent horizon increases monotonically with time, as shown in figure \ref{fig:5}, which is in accordance with the second law of black hole mechanics.

Interestingly, we find that the time required for the gravitational system to leave the initial RN-AdS black hole and enter the final scalarized black hole depends on the quench strength, satisfying $\tau\propto-\text{ln}|p|$ as shown in figure \ref{fig:6}.
The derivation of this relation is as follows.
The quench causes a deviation in the field configuration $\delta\phi$.
Due to the sufficiently small quench strength, the quenched gravitational system is stil in the linear region of the initial RN-AdS black hole, so that the perturbation can be linearly expanded based on the eigenmodes of the RN-AdS black hole, as follows
\begin{equation}
	\delta\phi_{p}(t,r)=f(p)e^{-iw_{*}t}\delta\phi_{*}(r)+\text{decaying modes},\label{eq:3.2}
\end{equation}
where $\delta\phi_{*}(r)$ is the only unstable eigenmode associated with the eigenvalue $w_{*}$ with a positive imaginary part.
Since the unstable mode is forever silent for $p=0$, the Taylor expansion of the coefficient funtion $f(p)$ lacks the constant term.
Assuming $f(p)\sim f_{n}p^{n}$, the sign that the system is far from equilibrium is that the coefficient of the unstable mode grows to a finite value
\begin{equation}
	f_{n}p^{n}e^{-iw_{*}\tau}\sim O(1).
\end{equation}
Taking the logarithm on both sides of the above equation, one can obtain
\begin{equation}
	\tau= -\frac{n}{\text{Im}[w_{*}]}\text{ln}|p|+C,\label{eq:3.4}
\end{equation}
where $C$ is a constant.
We observe that the slope of the straight line in figure \ref{fig:6} is reciprocal to that in the second stage in figure \ref{fig:4}, indicating $n=1$.
As a result, for a RN-AdS black hole with a single unstable mode, the time required for the scalarization satisfies the relation (\ref{eq:3.4}) with $n=1$ under the quench (\ref{eq:3.1}).

\subsection{Continuous descalarization}
From figure \ref{fig:1}, one can observe that the scalar condensation of the scalarized black hole decreases with the increase of the energy density.
Therefore, the key to descalarization is whether enough energy can be injected into the black hole.
Not surprisingly, it turns out that the spatial disturbance of the scalar field, which is converted into energy after being absorbed by the central black hole, is able to descalarize the scalarized black hole \cite{Zhang:2022cmu}.
A natural question is whether the holographic quench can inject enough energy into the scalarized black hole to trigger the descalarization.
To this end, the quench (\ref{eq:3.1}) is imposed on the scalarized black hole with energy density $\epsilon=1.6$.

\begin{figure}
	\begin{center}
		\subfigure[]{\includegraphics[width=.49\linewidth]{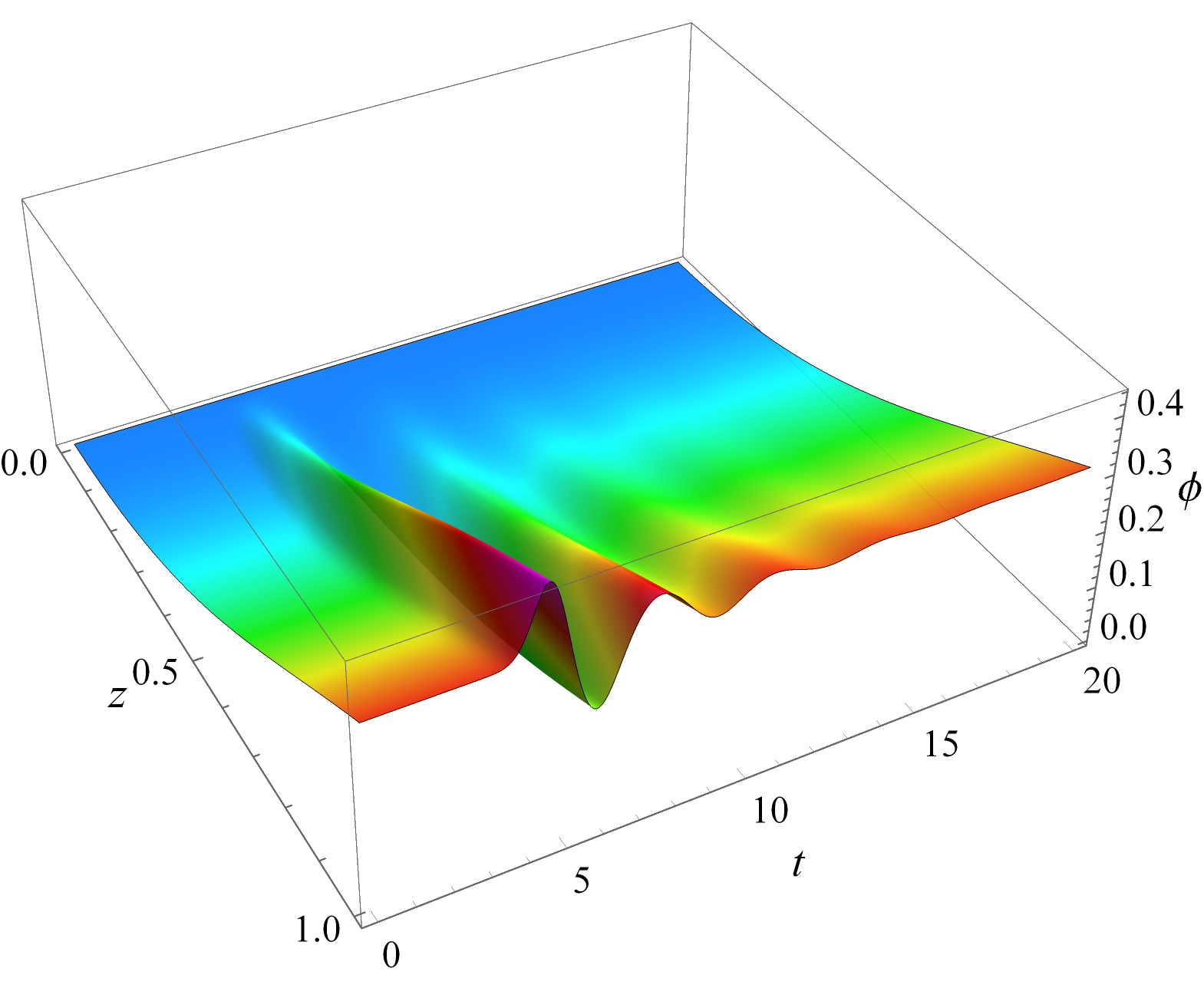}\label{fig:7}}
		\subfigure[]{\includegraphics[width=.49\linewidth]{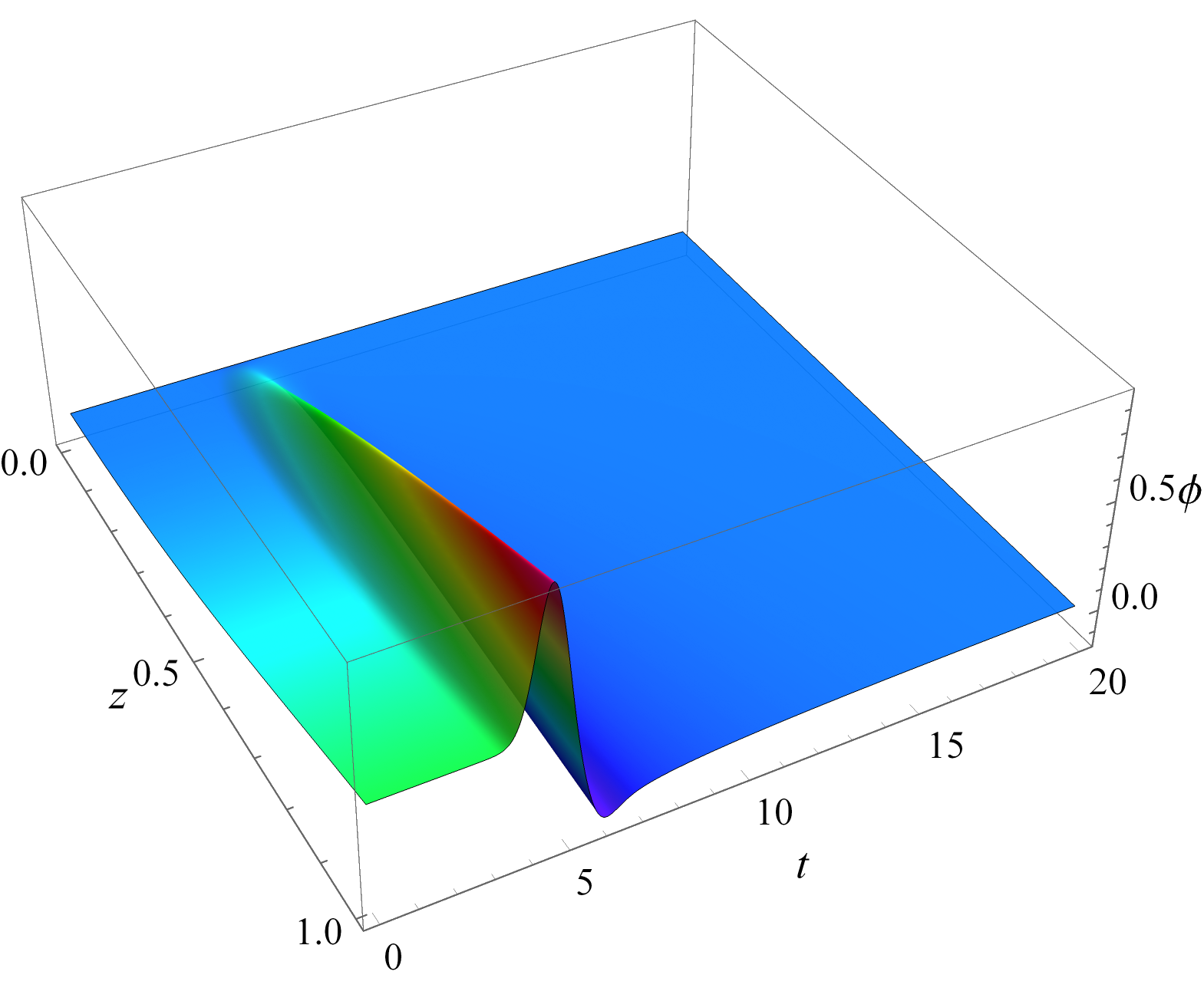}\label{fig:8}}
		\caption{The configuration of the scalar field as a function of time during the quench process with quench strength $p=0.2$ (a) and $p=1.2$ (b), respectively.
				The compactification coordinate $z=r^{-1}$ is introduced to constrain the radial direction to be finite.
				The apparent horizon and the AdS boundary are located at $z=1$ and $z=0$, respectively.}
		\label{fig:7-8}
	\end{center}
\end{figure}

The resulting real-time dynamics is shown in figure \ref{fig:7-8} as the evolution of the scalar field configuration over time.
One can observe that during the quench stage, the external source on the AdS boundary pulls the entire profile of the scalar field , rising with increasing source and falling with decreasing source.
In later stages of evolution, the dynamical behavior of the gravitational system depends on the specific quench strength.
For a weak quench with a strength of $p=0.2$ as shown in figure \ref{fig:7}, the scalar field accompanied by the oscillatory behavior gradually converges to a non-trivial configuration, leading to the formation of a scalarized black hole with more energy density but less scalar condensation.
On the other hand, for a strongth quench with a strength of $p=1.2$ as shown in figure \ref{fig:8}, the scalar field decays smoothly and exponentially, leaving a bald black hole with the vanishing scalar field.
At this time, the energy density of the gravitational system exceeds that represented by point $B$ in figure \ref{fig:1-2}, indicating that it enters the far-extremal region and can only reside on a RN-AdS black hole with dynamical stability.
As a result, the holographic quench mechanism in the EMs theory can inject enough energy into a black hole to trigger the transition from a scalarized black hole to a RN-AdS black hole, resulting in the descalarization phenomenon.

\begin{figure}
	\begin{center}
		\subfigure[]{\includegraphics[width=.49\linewidth]{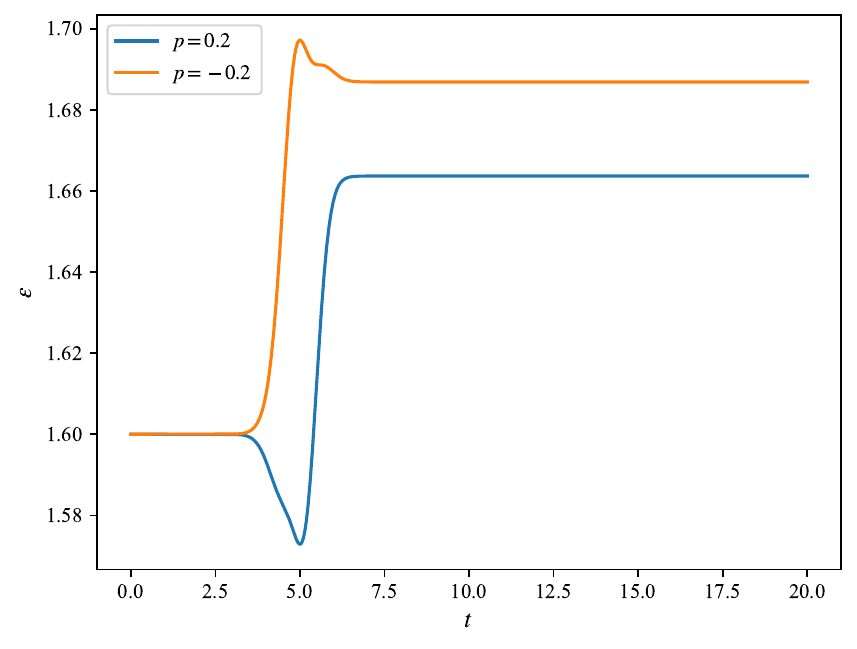}\label{fig:9}}
		\subfigure[]{\includegraphics[width=.49\linewidth]{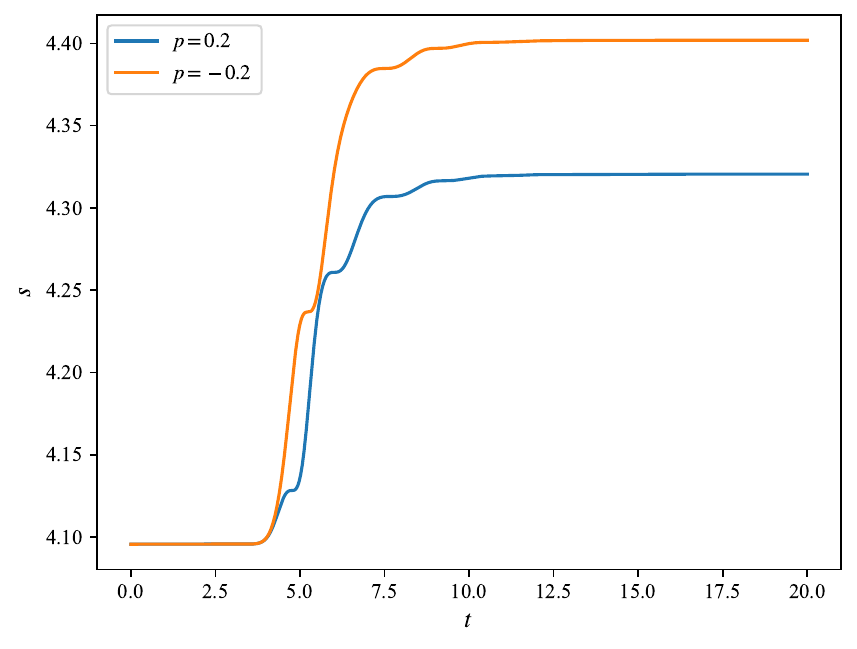}\label{fig:10}}
		\caption{The energy density (a) and entropy density (b) of the gravitational system as a function of time during the quench process.
				 The blue and orange lines represent the cases where the quench strength is $p=0.2$ and $p=-0.2$, respectively.}
		\label{fig:9-10}
	\end{center}
\end{figure}

In fact, during the quench stage, the process of injecting energy is not going well, manifested in the non-monotonic increase of the energy density as shown in figure \ref{fig:9}.
From the Ward-Takahashi identity (\ref{eq:2.18}), the trend of the evolution of the energy density is reversed as the sign of the quench amplitude changes.
However, the appearance of the final states with more energy in figure \ref{fig:9} indicates that the quench with an amplitude of either positive or negative sign corresponds to the process of injecting energy.
We have also checked other forms of quench that require both initial and final values to be zero but fail to find the dynamical process for energy extraction from AdS spacetime.
This is different from the case of quenching a charged scalar field \cite{Chen:2022vag}, where the process of energy extraction can be triggered by the superradiance \cite{Ishii:2022lwc}, which in our case is suppressed by the real scalar field.
Although the energy density can increase or decrease during the quench process, depending on the time dependence of the scalar source, the area of the apparent horizon always increases monotonically with time, indicating that the second law of black hole mechanics holds even in the case of the time-dependent scalar source, as shown in figure \ref{fig:10}.

\begin{figure}
	\begin{center}
		\subfigure[]{\includegraphics[width=.49\linewidth]{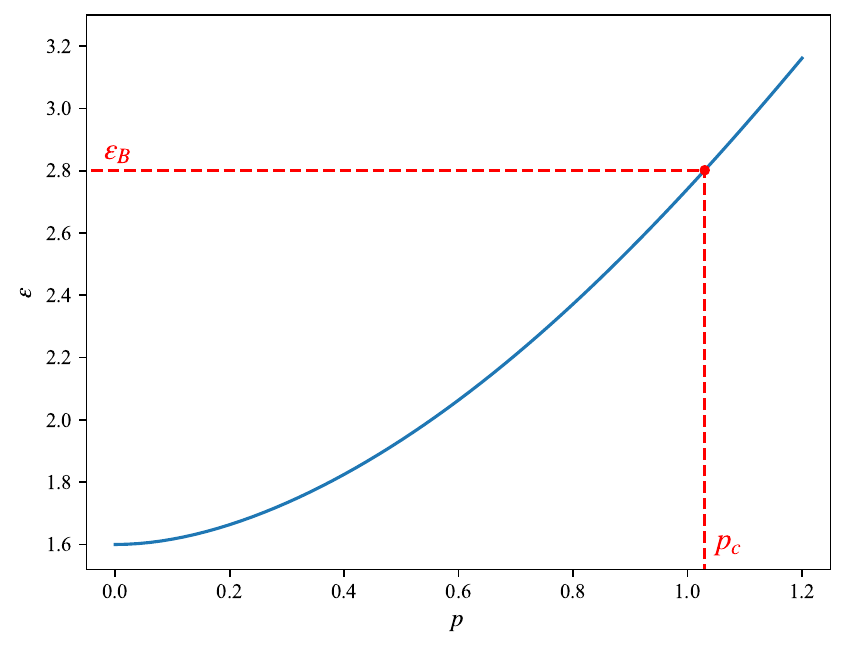}\label{fig:11}}
		\subfigure[]{\includegraphics[width=.49\linewidth]{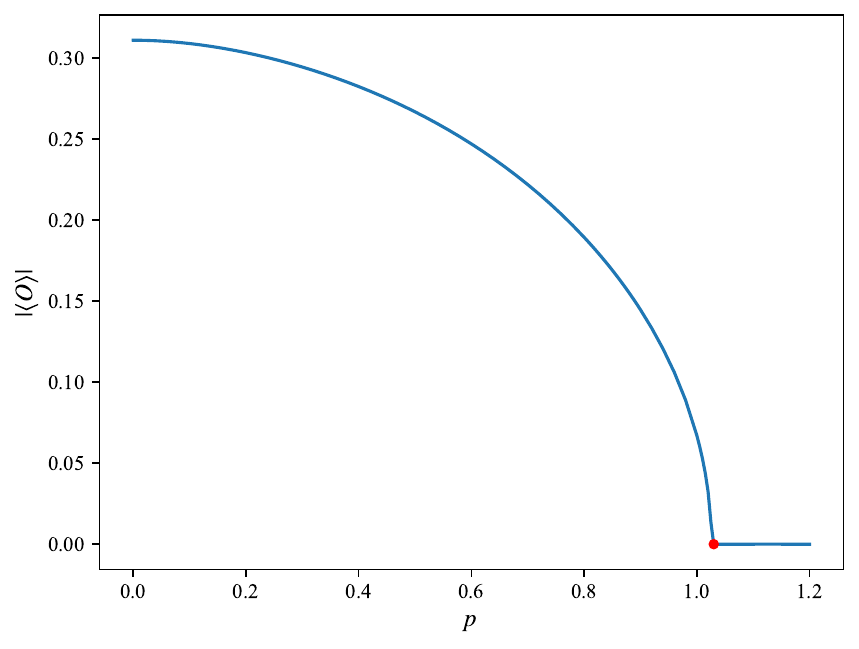}\label{fig:12}}
		\caption{The energy density (a) and the expectation value of the scalar operator (b) of the final state after quenching as a function of quench strength. 
				The red point represents the critical quench strength $p_{c}$ required for the transition from the initial scalarized black hole to the RN-AdS black hole with the energy density represented by point $B$ in figure \ref{fig:1-2}.}
		\label{fig:11-12}
	\end{center}
\end{figure}

In figure \ref{fig:11-12}, we show that the energy density and the expectation value of the scalar operator for the final state of the quenched system vary with the quench strength.
One can observe that the energy density of the gravitational system increases monotonically with the increase of the quench strength.
There exists a critical quench strength $p_{c}$ corresponding to the energy density represented by point $B$ in figure \ref{fig:1-2}, separating the scalarized black holes and RN-AdS black holes.
For the subcritical strengths, the quench smoothly reduces the scalar condensation carried by the initial scalarized black hole without changing its essential properties.
On the other hand, the quench with a supercritical strength will trigger a transition, descalarizing the initial scalarized black hole and resulting in the emergence of a RN-AdS black hole residing in the far-extremal region.
In addition to the mechanism of injecting energy, the descalarization phenomenon can also be achieved by quenching a charged scalar field, which induces a loss of charge \cite{Chen:2022vag}.
Different from the continuous descalarization here, the oscillatory behavior of the charge with the quench strength leads to the appearance of the repeated scalarization and repeated descalarization phenomena.
For a sufficiently large quench strength, the charge of the gravitational system is completely lost, leaving a Schwarzschild-AdS black hole instead of the RN-AdS black hole with more energy.

\subsection{Dynamical phase diagram}

\begin{figure}
	\begin{center}
		\subfigure[]{\includegraphics[width=.49\linewidth]{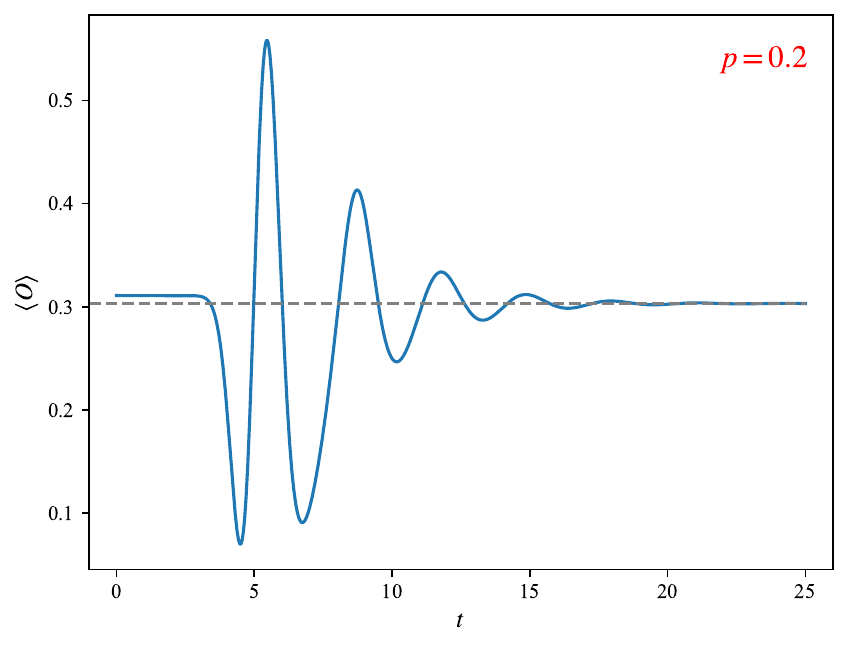}\label{fig:13}}
		\subfigure[]{\includegraphics[width=.49\linewidth]{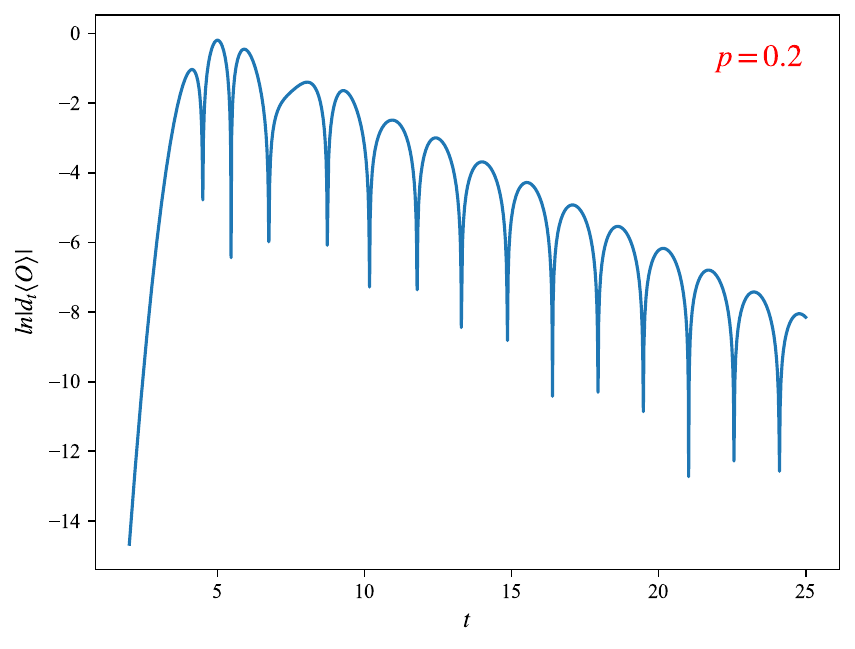}\label{fig:14}}
		\subfigure[]{\includegraphics[width=.49\linewidth]{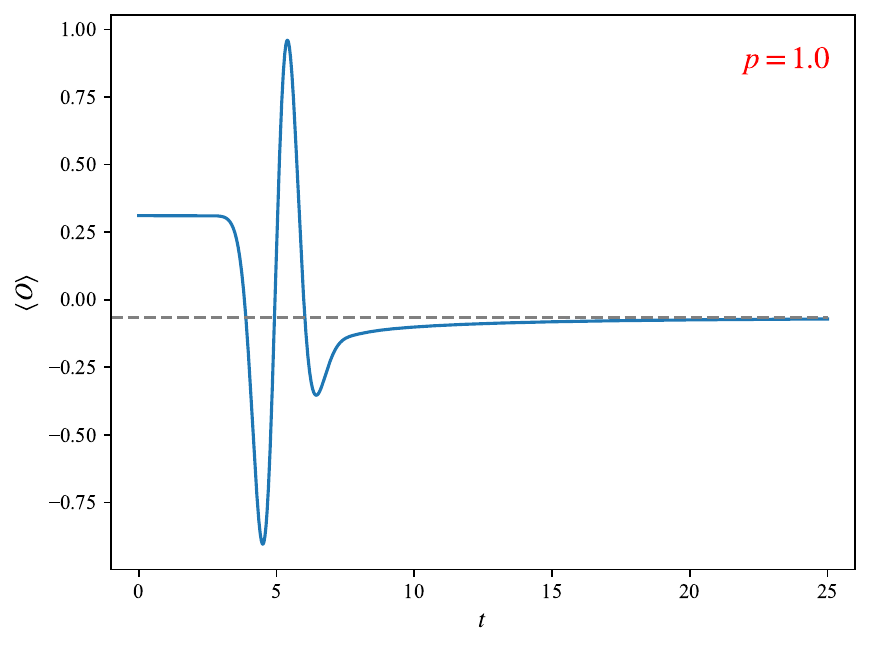}\label{fig:15}}
		\subfigure[]{\includegraphics[width=.49\linewidth]{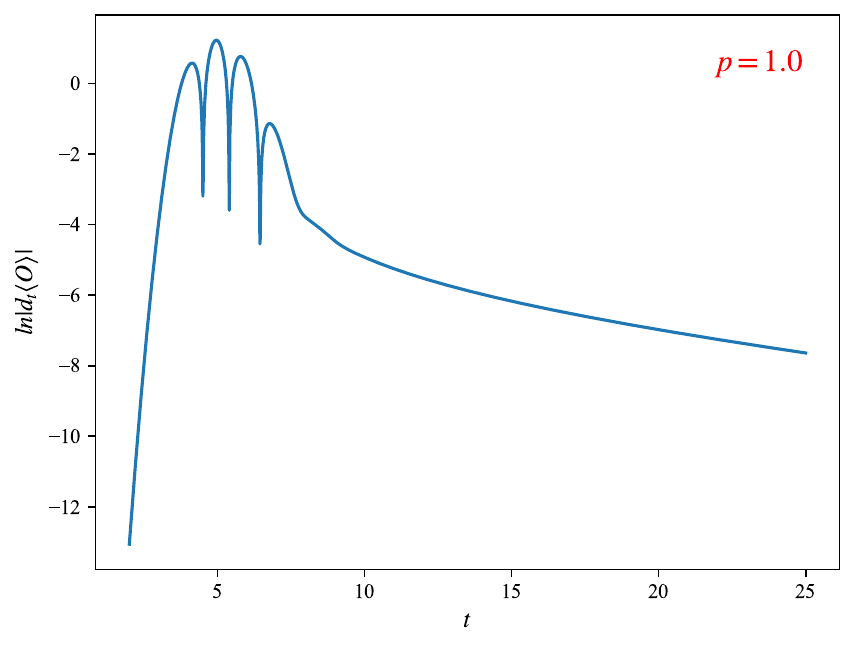}\label{fig:16}}
		\subfigure[]{\includegraphics[width=.49\linewidth]{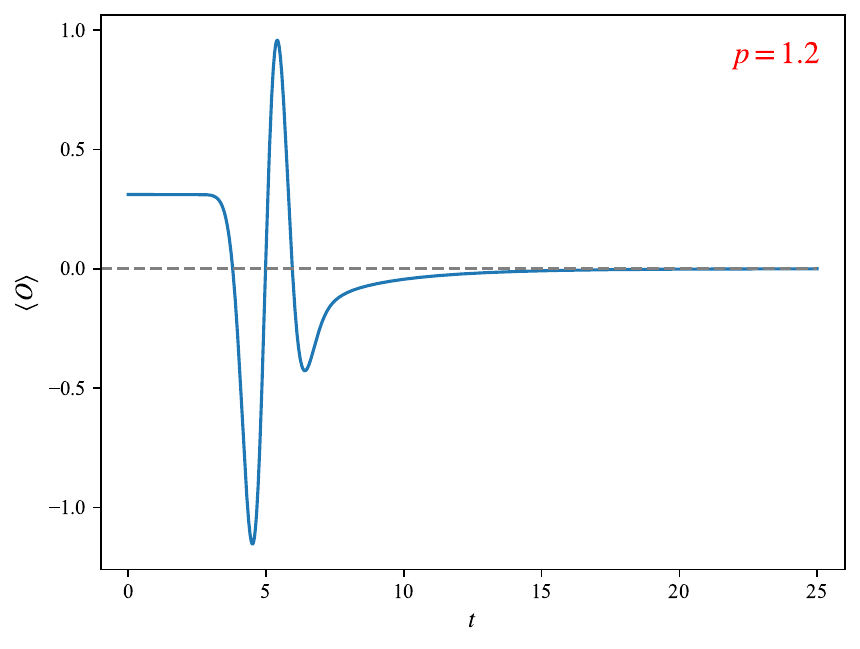}\label{fig:17}}
		\subfigure[]{\includegraphics[width=.49\linewidth]{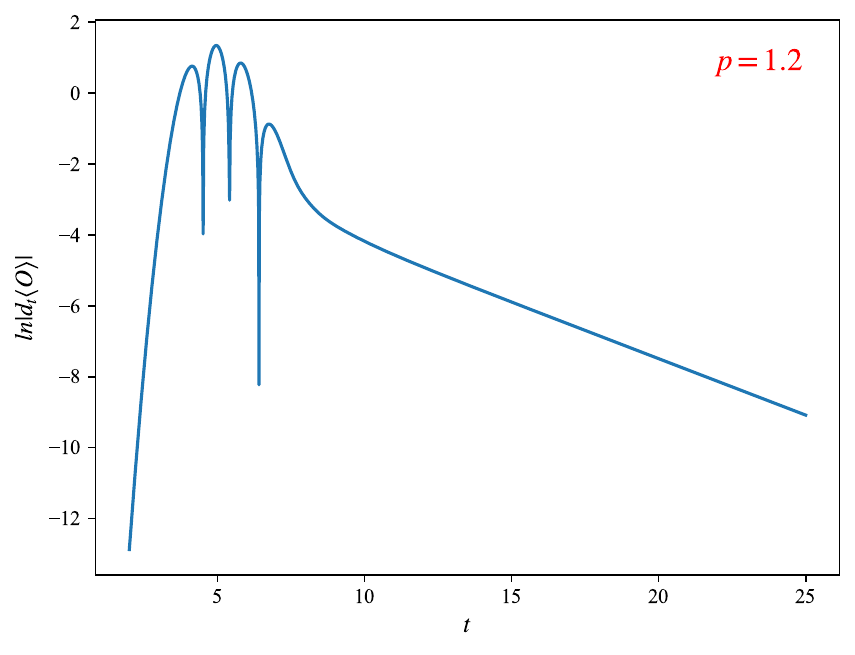}\label{fig:18}}
		\caption{The expectation value of the scalar operator (left panels) and the value of ln$|d_{t}\left\langle O\right\rangle|$ (right panels) as a function of time for different quench strengths.
				The dynamics of later times exhibits three types.
				For $p=0.2$, one can observe an oscillatory decay towards $\left\langle O\right\rangle\neq 0$.
				For $p=1.0$, the oscillatory behavior disappears and the scalar condensation decays directly towards $\left\langle O\right\rangle\neq 0$.
				For $p=1.2$, it converges towards $\left\langle O\right\rangle= 0$ without oscillation.}
		\label{fig:13-18}
	\end{center}
\end{figure}

From figure \ref{fig:7-8}, one can observe that the dynamical behavior of the scalar condensation at late times shows a significant difference, oscillating or not, for the two cases where the final state is a scalarized black hole or a RN-AdS black hole.
Actually, before the transition from the scalarized black hole to a RN-AdS black hole occurs, such a dynamical transition from oscillating to non-oscillating
has already appeared.
As shown in figure \ref{fig:13-18}, the real-time dynamics of the quenched scalarized black hole can be divided into three cases.
For a quench with a weaker strength, the scalar condensation decays exponentially to a slightly smaller constant at late times, accompanied by a strongly oscillatory behavior.
The period of such oscillation becomes longer as the quench strength increases.
Until the quench strength exceeds a critical value $p_{*}$, which is smaller than $p_{c}$, the oscillatory behavior in the late evolution completely disappears.
The scalar condensation decays exponentially to a non-trivial value directly.
As the quench strength continues to increase to approache $p_{c}$, the expectation value of the scalar operator of the final state gradually converges to zero.
After that, the gravitational system always settles down to a RN-AdS black hole without oscillation in the late stage.

\begin{figure}
	\begin{center}
		\subfigure[]{\includegraphics[width=.49\linewidth]{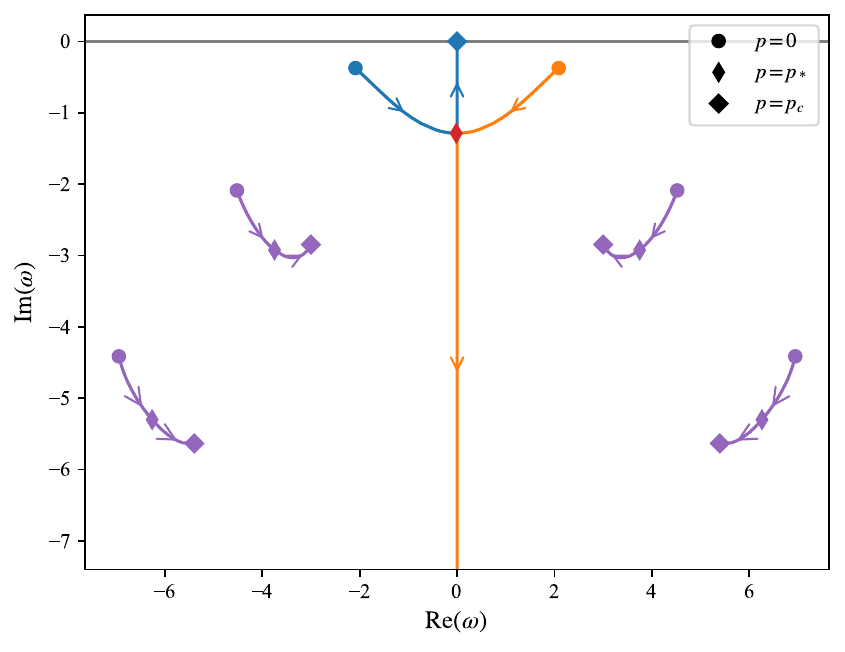}\label{fig:19}}
		\subfigure[]{\includegraphics[width=.49\linewidth]{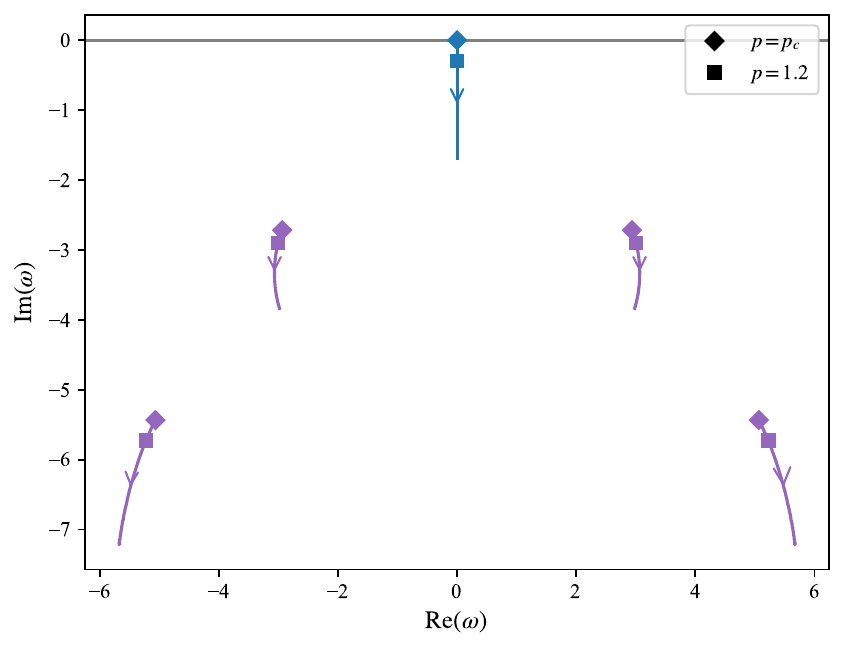}\label{fig:20}}
		\subfigure[]{\includegraphics[width=.49\linewidth]{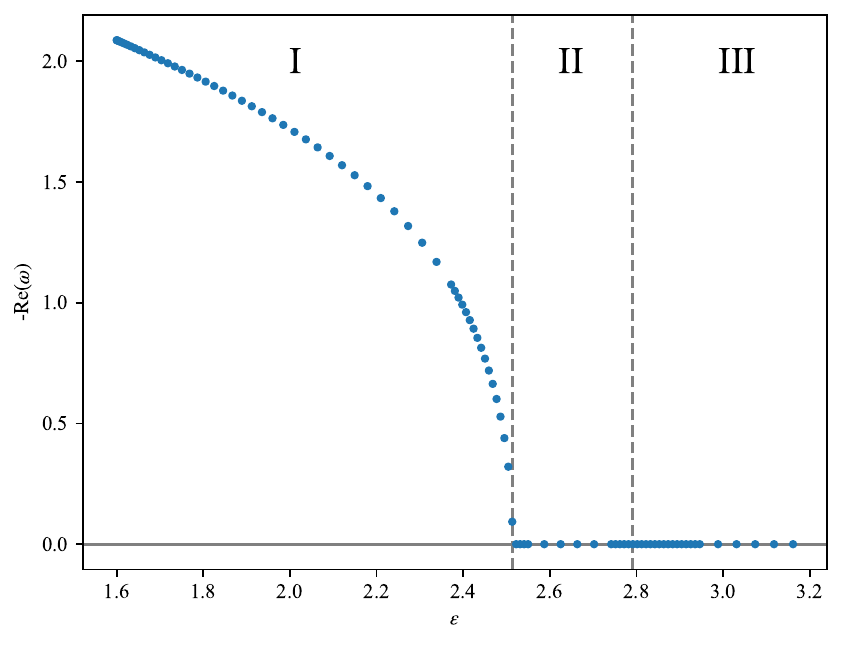}\label{fig:21}}
		\subfigure[]{\includegraphics[width=.49\linewidth]{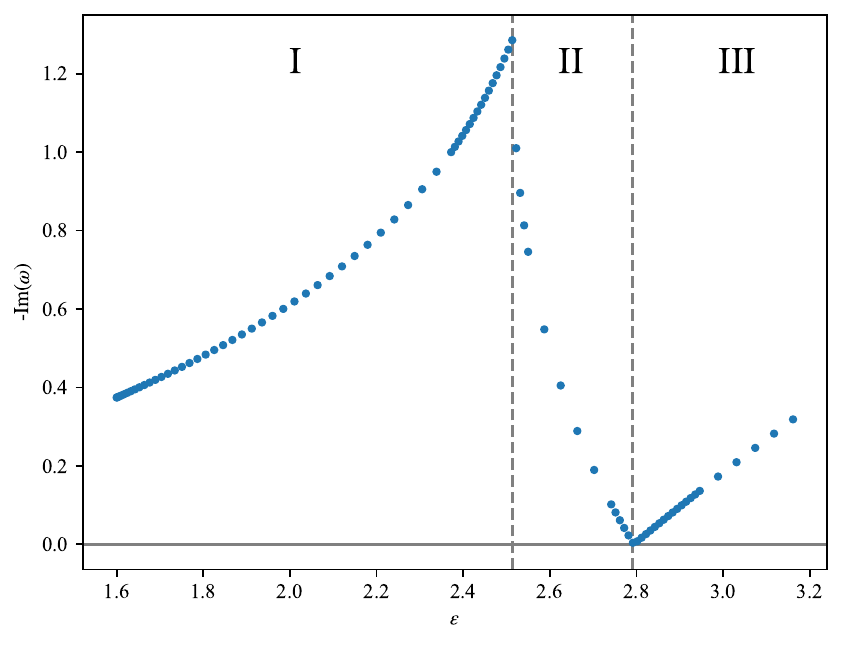}\label{fig:22}}
		\caption{The trajectories of the quasi-normal modes of the final states, scalarized black holes (a) and RN-AdS black holes (b), with respect to the quench strength.
			On each trajectory, the arrow indicates the direction of increasing energy density.
			The blue and orange dots represent the dominant modes, and the purple dots represent other modes.
			(c, d) The real and imaginary parts of the dominant mode represented by the blue dot as a function of energy density.}
		\label{fig:19-22}
	\end{center}
\end{figure}

Such an oscillatory behavior depends on the dominant quasi-normal mode of the final state.
In the later stage of evolution, the gravitational system enters the linear region of the dynamically stable final state, which can be approximated as
\begin{equation}
	\left\langle O(t)\right\rangle\approx\left\langle O\right\rangle_{f}+\mathcal{A}e^{-iwt},\label{eq:3.5}
\end{equation}
where $\left\langle O\right\rangle_{f}$ and $w$ represent the expectation value of the scalar operator and the eigenvalue of the dominant mode of the final state, respectively. 
$\mathcal{A}$ is an amplitude prefactor.
Due to the real configuration of the scalar field, the quasi-normal spectrums necessarily possess the following symmetry
\begin{equation}
	w\rightarrow-w^{*},\label{eq:3.6}
\end{equation}
as shown in figure \ref{fig:19-22}. 
For the dynamically stable modes with a negative imaginary part, the real part characterizes the oscillation and the imaginary part corresponds to decay.
Among them, the mode with the smallest absolute value of the imaginary part has the slowest decay rate, which is called the dominant mode.
As shown in figure \ref{fig:19}, for the scalarized black holes, as the quench strength gradually increases and reaches $p_{*}$, the dominant mode pair migrates towards the imaginary axis and finally meet on it, indicating the disappearing real part.
Then as the quench strength approaches from $p_{*}$ to $p_{c}$, one of them climbs up along the imaginary axis to approach the origin, while the other moves down sharply.
At $p=p_{c}$, the zero mode appears, accompanied by the transition to a RN-AdS black hole.
Subsequently, the dominant mode shifts down along the imaginary axis with the increase of the quench strength, as shown in figure \ref{fig:20}.

That is to say, there are two critical quench strength to distinguish three regimes of non-equilibrium response, as shown in figures \ref{fig:21} and \ref{fig:22}.
The transition from I to II is marked by the fact that the real part of the dominant mode reaches zero, accompanied by a turning point in the imaginary part, similar to the case of holographic superfluids under quantum quenches \cite{Bhaseen:2012gg} or even periodic driving \cite{Li:2013fhw}.
In region I, the scalar condensation exhibits oscillatory exponential decay with Re$(w)\neq 0$ towards $\left\langle O\right\rangle_{f}\neq 0$, with the decay rate that increases with the quench strength.
However, in region II, the oscillatory behavior disappears due to Re$(w)= 0$, and the scalar condensation decays exponentially towards $\left\langle O\right\rangle_{f}\neq 0$ at a rate that decreases with the increasing quench strength.
On the other hand, the transition from II to III occurs when the imaginary part of the dominant mode reaches zero, accompanied by the disappearance of the scalar condensation.
Different from the case of holographic superfluid, where the real part of the dominant mode in region III is non-zero, in our case the dominant mode always possesses a zero real part.
Nonetheless, substituting $\left\langle O\right\rangle_{f}=0$ into (\ref{eq:3.5}), in both cases, the absolute value of the expectation value exhibits a similar non-oscillatory exponential decay behavior in the dynamical process.
As a result, from the dynamics of the boundary scalar operator after a thermal quench, a scalarized black hole in the EMs theory behaves like a holographic superfluid.

\begin{figure}
	\begin{center}
		\subfigure[]{\includegraphics[width=.49\linewidth]{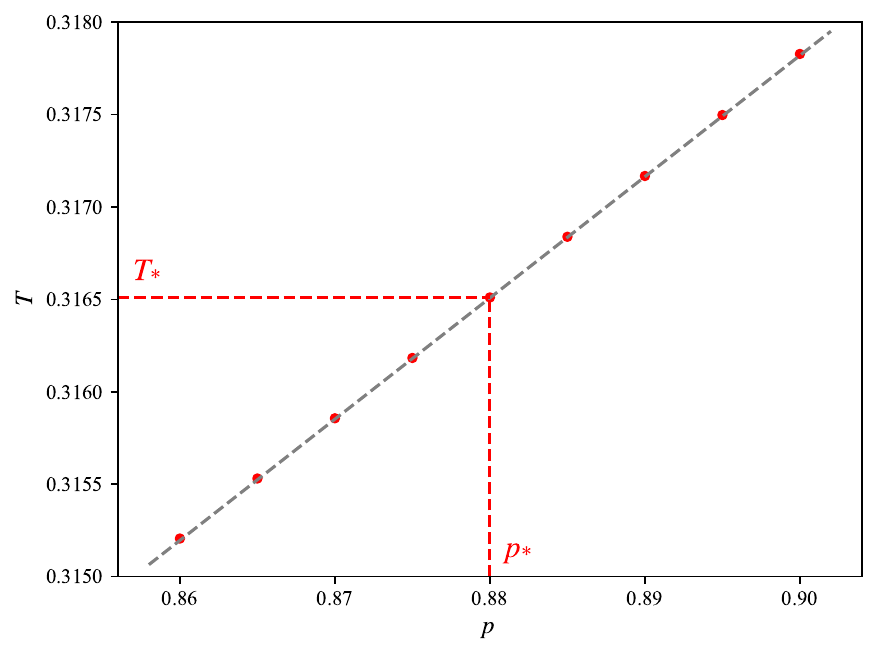}\label{fig:23}}
		\subfigure[]{\includegraphics[width=.49\linewidth]{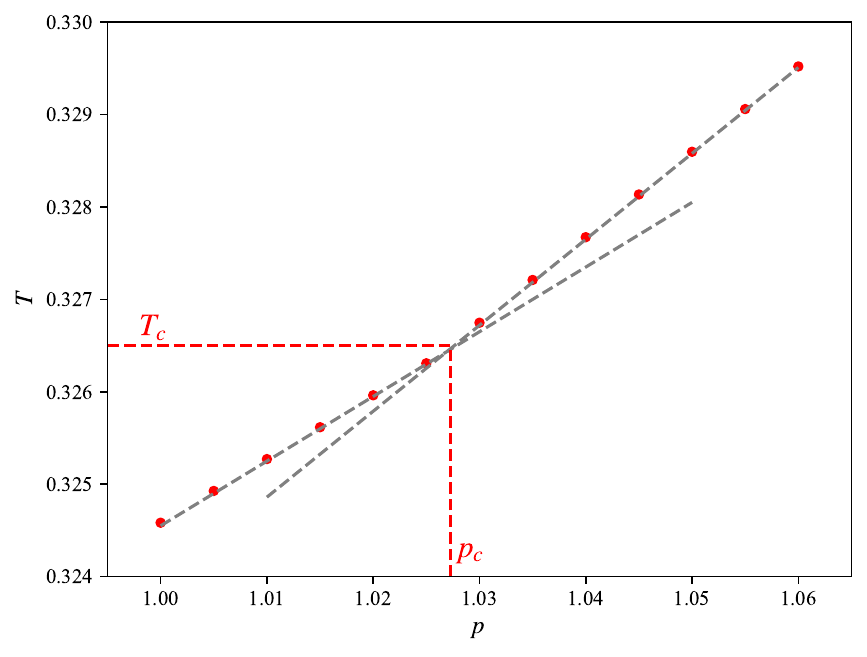}\label{fig:24}}
		\caption{The temperature of the final state as a function of quench strength near $p_{*}$ (a) and $p_{c}$ (b).}
		\label{fig:23-24}
	\end{center}
\end{figure}

In figure \ref{fig:23-24}, we show that the temperature of the final state in the vicinity of the two transitions varies with the quench strength.
One can observe that the temperature increases monotonically with the quench strength in both cases.
Different from the transition from I to II, where the temperature grows linearly with almost the same rate, in the case of transition from II to III there is a visible turning point in the change in temperature.

\section{Critical behaviors}\label{sec:Cb}
By continuously increasing the quench strength, we have found two critical strengths to trigger two different dynamical transitions during the descalarization process.
Actually, there exists another critical quench strength that induces a class of critical dynamical behaviors.
With the coupling function (\ref{eq:coupling}), the action (\ref{eq:action}) is $\mathbb Z_{2}$-invariant under the transformation $\phi\rightarrow-\phi$.
Such a $\mathbb Z_{2}$-symmetry indicates that for a gravitational system with the energy density in the $AB$ region in figure \ref{fig:1-2}, in addition to the dynamically unstable RN-AdS black hole in the excited state, there are two degenerate ground states, the scalarized black holes with positive and negative scalar condensations, respectively.
The spontaneous scalarization simulated in subsection \ref{sec:ss} describes the spontaneous transition process from the excited state to a ground state.
The choice of the final state depends on the specific sign of the quench (\ref{eq:3.1}).
For a small quench strength, the positive amplitude corresponds to the ground state with positive scalar condensation, whereas the negative one causes the system to fall into the other.
Of course, such two time-dependent dynamical processes are degenerate, with a difference of a negative sign overall.
A natural question is whether there exists a dynamical path connecting the two degenerate ground states.
From figure \ref{fig:15}, one can observe that the quench of sufficient strength can excite the transition of the gravitational system from one ground state to the other, whereas the quench of weak strength cannot, as shown in figure \ref{fig:13}.
Such results suggest the existence of a dynamical barrier in this transition, manifested as a threshold of the quench strength.
It is illuminating to consider the dynamical behaviors of the gravitational system near the threshold.

\begin{figure}
	\begin{center}
		\subfigure[]{\includegraphics[width=.49\linewidth]{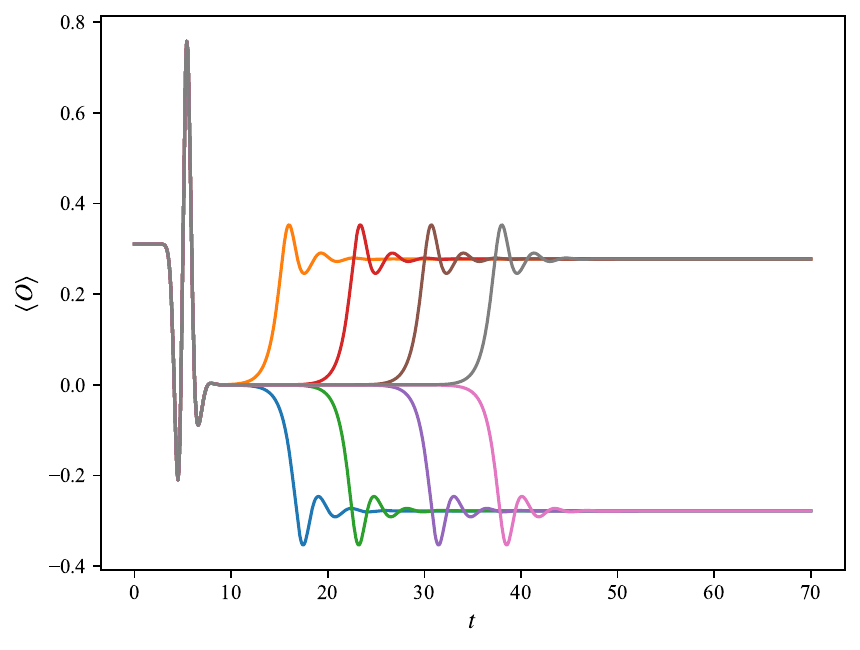}\label{fig:25}}
		\subfigure[]{\includegraphics[width=.49\linewidth]{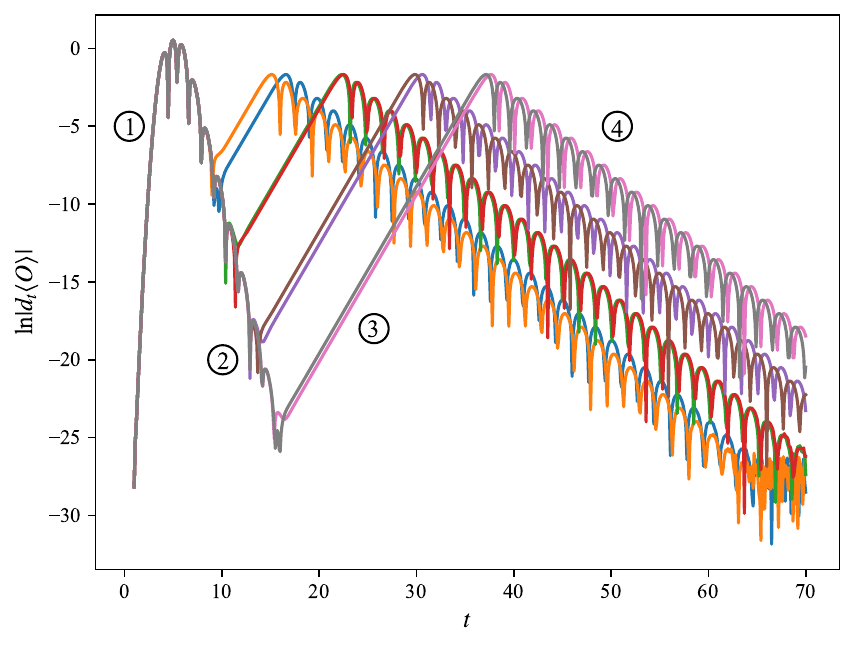}\label{fig:26}}
		\subfigure[]{\includegraphics[width=.49\linewidth]{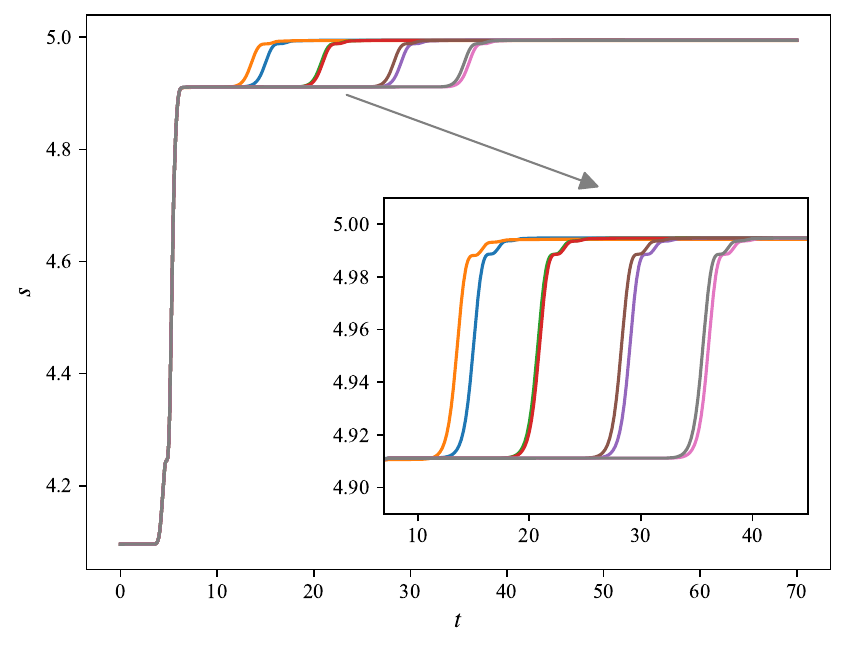}\label{fig:27}}
		\subfigure[]{\includegraphics[width=.49\linewidth]{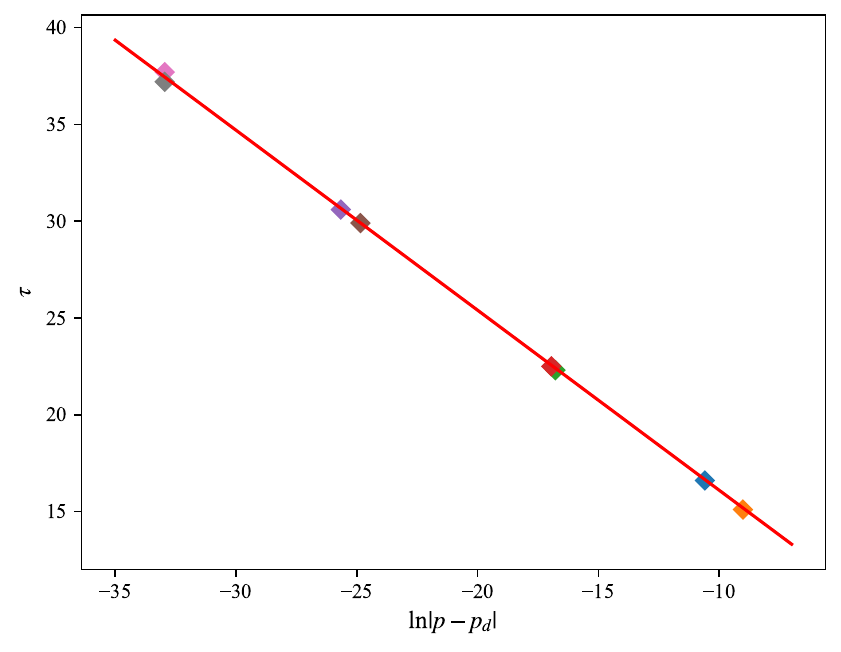}\label{fig:28}}
		\caption{The expectation value of the scalar operator (a), the value of ln$|d_{t}\left\langle O\right\rangle|$ (b) and the entropy density (c) as a function of time during the descalarization process. 
			The quench strength $p$ is close to the threshold $p_{d}$. 
			(d):  The time $\tau$ that the quenched system stays near the critical RN-AdS black hole with respect to ln$|p-p_{d}|$.
			All the curves and points of the same color in the figure correspond to each other.}\label{fig:25-28}
	\end{center}
\end{figure}

Continuously approaching the threshold $p_{d}$ through the dichotomy, the resulting real-time dynamics is shown in figure \ref{fig:25-28}.
Interestingly, after the quench stage, the gravitational system is attracted to the vicinity of a critical RN-AdS black hole with dynamical instability, manifested by the complete disappearance of the scalar condensation in the dynamical intermediate process for a period of time, as shown in figure \ref{fig:25}.
At the same time, the area of the apparent horizon stops growing after the drastic increase caused by the quench, leading to the appearance of a plateau, as shown in figure \ref{fig:27}.
From figure \ref{fig:26}, one can observe that during the time the gravitational system stays near the critical RN-AdS black hole, it goes through two dynamical processes, represented as the second and third stages in the figure.
Firstly, in the second stage, the evolving system is dominated by the stable modes of the critical RN-AdS black hole, exhibiting an exponential decay behavior.
After that, the single unstable mode of the critical RN-AdS black hole is excited in the third stage, pushing the system exponentially away from the critical state into the linear region of the final stable scalarized black hole.
Due to the drastic change of the gravitational configuration, the area of the apparent horizon also starts to grow synchronously.
As we mentioned earlier, for a scalarized black hole, there are two degenerate states.
We find that the quench with a subcritical strength fails to trigger the dynamical transition, corresponding to a scalarized black hole with a positive scalar condensation as the initial value.
Conversely, the quench with a supercritical strength induces the system to evolve into a scalarized black hole with a negative scalar condensation.
In addition, the time for the gravitational system to stay on the critical RN-AdS black hole in the dynamical intermediate process depends on the gap between the actual quench strength and the threshold value, satisfying (\ref{eq:3.4}), as shown in figure \ref{fig:28}, where the slope of the line is equal to the reciprocal of the imaginary part of the single unstable mode of the critical RN-AdS black hole.
From this it can be inferred that the threshold $p_{d}$ corresponds exactly to the critical RN-AdS black hole with dynamical instability.
As a result, in the EMs theory, the holographic quench mechanism can induce the dynamical transition of the gravitational system with a specific energy bidirectionlly between two degenerate ground states by crossing an excited state.

\begin{figure}
	\begin{center}
		\subfigure[]{\includegraphics[width=.49\linewidth]{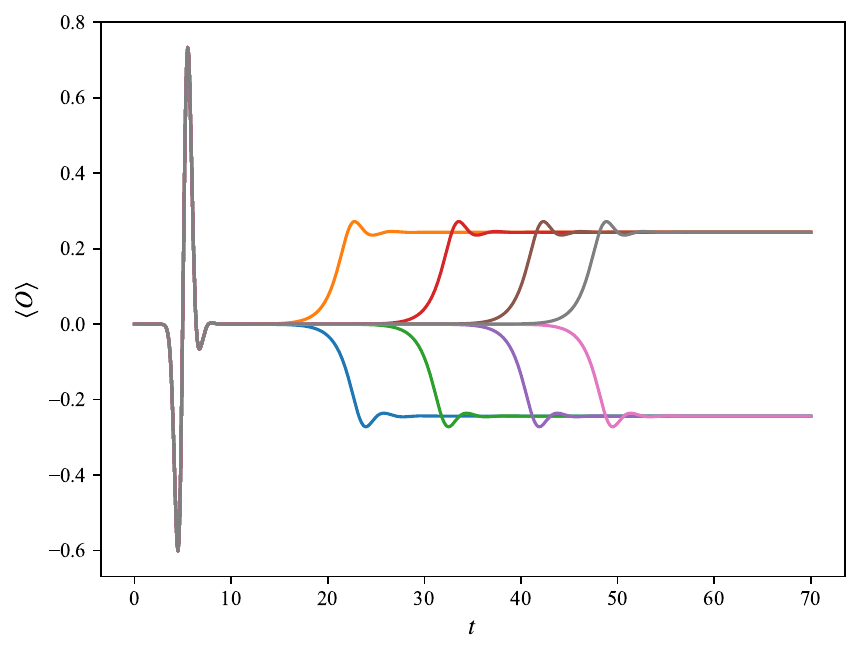}\label{fig:29}}
		\subfigure[]{\includegraphics[width=.49\linewidth]{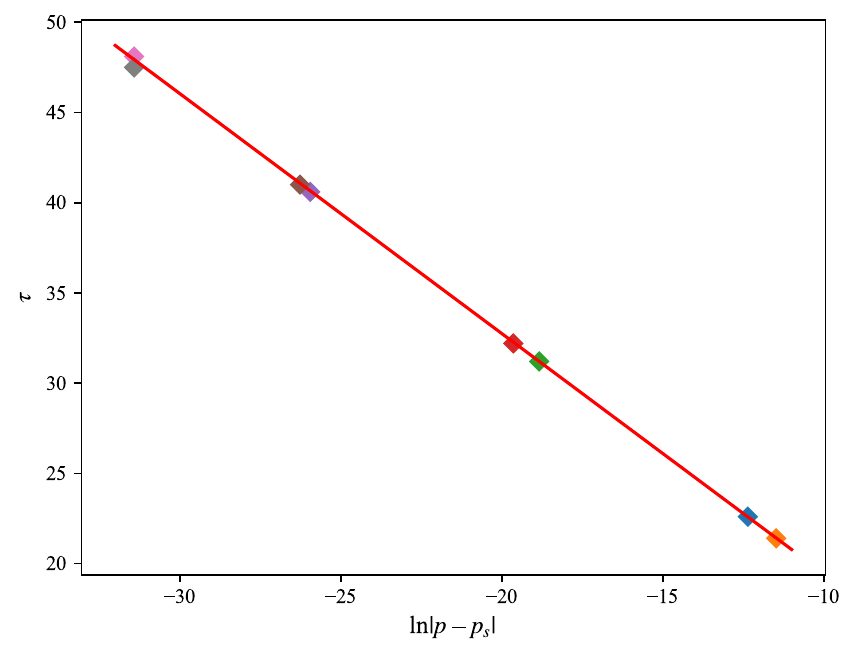}\label{fig:30}}
		\caption{(a): The expectation value of the scalar operator as a function of time during the scalarization process.
			The quench strength $p$ is close to the threshold $p_{s}$. 
			(b):  The time $\tau$ that the quenched system stays near the critical RN-AdS black hole with respect to ln$|p-p_{s}|$.
			All the curves and points of the same color in the figure correspond to each other.}\label{fig:29-30}
	\end{center}
\end{figure}

Unexpectedly, a similar dynamical critical phenomenon also appears in the spontaneous scalarization process with a dynamical unstable RN-AdS black hole as the initial value, as shown in figure \ref{fig:29-30}.
There exists a non-zero threshold $p_{s}$ separating the two degenerate scalarized black holes, near which the evolving system is attracted to another dynamical unstable RN-AdS black hole with more energy in the dynamical intermediate process.
That is to say, the quench (\ref{eq:3.1}) with the strength $p_{s}$ fails to trigger the dynamical instability of the initial RN-AdS black hole, injecting only energy into it.
Such results indicate that the critical dynamical behavior is quite universal to the initial value.

\section{Conclusion}\label{sec:C}
In the EMs theory with a quadratic coupling, the domain of existence of solutions is composed of a branch of RN-AdS black holes and a branch of degenerate scalarized black holes.
In the energy range where these two kinds of black holes coexist, the dynamically unstable RN-AdS black hole is in the excited state, and the two degenerate, thermodynamically favored scalarized black holes are in the two ground states.
After a quench, a series of interesting dynamical transitions will occur in them.

For the excited state, the dynamically unstable RN-AdS black hole will be scalarized spontaneously and evolve into a ground state under perturbations.
In particular, in the weak quench regime, the time for the scalarization of a RN-AdS black hole with a single unstable mode satisfies (\ref{eq:3.4}).
When the quench strength continues to increase beyond a threshold $p_{c}$, the final fate of the scalarization transitions to the other degenerate ground state.
Near the threshold, a critical RN-AdS black hole with more energy emerges in the dynamical intermediate process.

\begin{figure}
	\begin{center}
		\includegraphics[width=.49\linewidth]{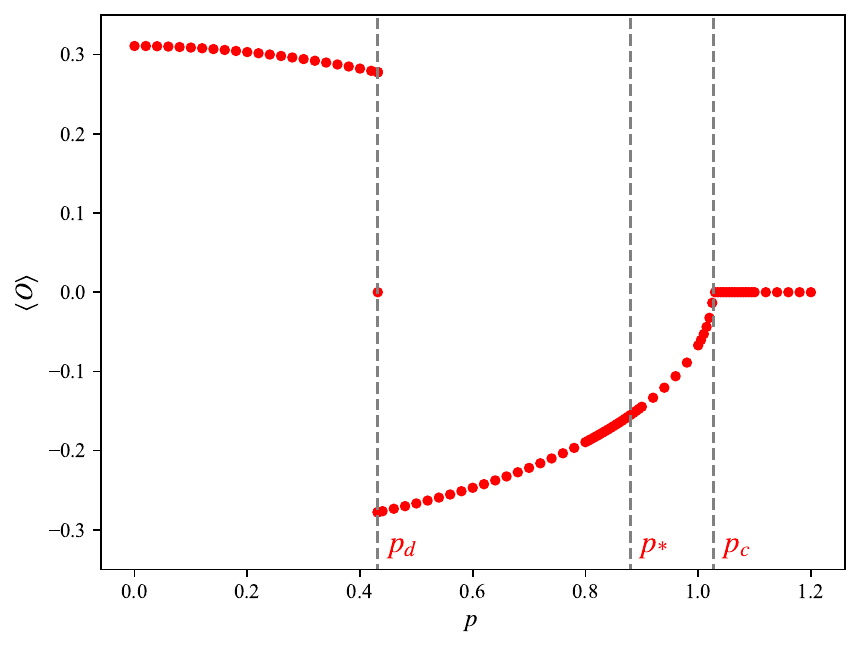}
		\caption{Dynamical phase diagram of the scalarized black hole showing the expectation value of the scalar operator as a function of quench strength.}
		\label{fig:31}
	\end{center}
\end{figure}

On the other hand, for a scalarized black hole, there are three critical quench strengths that trigger three different dynamical transitions, as shown in figure \ref{fig:31}.
The first critical strength $p_{d}$ induces the gravitational system to cross a critical RN-AdS black hole acting as a dynamical barrier and transition from one ground state to the other degenerate ground state.
Near the threshold $p_{d}$, the system is attracted to the excited state and stays there for a time dependent on the gap between the actual quench strength and the threshold $p_{d}$.
Such a critical dynamical phenomenon is induced by the $\mathbb Z_{2}$-symmetry of the model.
Near the second critical strength $p_{*}$, a dynamical transition similar to the case of holographic superfluid occurs, manifested by the disappearance of the oscillatory behavior during the decay to the final scalarized black hole.
From the perspective of quasi-normal modes, this occurs when the dominant mode of the final state migrates to the imaginary axis with increasing quench strength.
Until the quench strength exceeds the third critical value $p_{c}$, the injection of sufficient energy makes the energy density of the gravitational system reach the range where only the RN-AdS black hole exists, resulting in the descalarization phenomenon, analogues to the holographic superfluid phase transition from a superfluid state to a normal state.

A natural direction in the future is to further investigate the holographic description of the EMs theory.
At present, the holographic dual for the hairy black holes with a real scalar field is mostly confined to the EM-dilaton models, from the holographic QCD \cite{DeWolfe:2010he,DeWolfe:2011ts,Cai:2012xh,He:2013qq,Rougemont:2015ona,Knaute:2017lll,Giataganas:2017koz,Cai:2022omk} to the holographic superfluids and superconductors \cite{Aprile:2009ai,Liu:2010ka,Goldstein:2010aw,Cadoni:2011kv,Gouteraux:2011qh,Salvio:2012at}. 
The results of this paper reveal the potential connection between the EMs scalarized black holes and holographic superfluids, which to a certain extent indicates the holographic applicability of the EMs models.
Another topic worthy of investigation is the quench dynamics in the EMs theory with higher order coupling, in which there are two non-degenerate local ground states and one excited state \cite{Blazquez-Salcedo:2020nhs,LuisBlazquez-Salcedo:2020rqp,Chen:2023eru}.
The quench mechanism is expected to be able to overcome the dynamical barrier to realize the bidirectional dynamical transition between the two local ground states.

\appendix

\section*{Acknowledgement}
This work is supported in part by the National Natural Science Foundation of China under Grant Nos. 11975235, 12005077, 12035016, 12075026, 12275350, the National Key Research and Development Program of China Grant No. 2021YFC2203001 and the Guangdong Basic and Applied Basic Research Foundation under Grant No. 2021A1515012374. 

\bibliographystyle{JHEP}
\bibliography{references}

\providecommand{\href}[2]{#2}\begingroup\raggedright\begin{thebibliography}{10}

\bibitem{Hartnoll:2008vx}
S.A.~Hartnoll, C.P.~Herzog and G.T.~Horowitz, \emph{{Building a Holographic
  Superconductor}},
  \href{https://doi.org/10.1103/PhysRevLett.101.031601}{\emph{Phys. Rev. Lett.}
  {\bfseries 101} (2008) 031601}
  [\href{https://arxiv.org/abs/0803.3295}{{\ttfamily 0803.3295}}].

\bibitem{Hartnoll:2008kx}
S.A.~Hartnoll, C.P.~Herzog and G.T.~Horowitz, \emph{{Holographic
  Superconductors}},
  \href{https://doi.org/10.1088/1126-6708/2008/12/015}{\emph{JHEP} {\bfseries
  12} (2008) 015} [\href{https://arxiv.org/abs/0810.1563}{{\ttfamily
  0810.1563}}].

\bibitem{Herzog:2008he}
C.P.~Herzog, P.K.~Kovtun and D.T.~Son, \emph{{Holographic model of
  superfluidity}},
  \href{https://doi.org/10.1103/PhysRevD.79.066002}{\emph{Phys. Rev. D}
  {\bfseries 79} (2009) 066002}
  [\href{https://arxiv.org/abs/0809.4870}{{\ttfamily 0809.4870}}].

\bibitem{Astefanesei:2019pfq}
D.~Astefanesei, C.~Herdeiro, A.~Pombo and E.~Radu,
  \emph{{Einstein-Maxwell-scalar black holes: classes of solutions, dyons and
  extremality}}, \href{https://doi.org/10.1007/JHEP10(2019)078}{\emph{JHEP}
  {\bfseries 10} (2019) 078}
  [\href{https://arxiv.org/abs/1905.08304}{{\ttfamily 1905.08304}}].

\bibitem{Kaluza:2018}
T.~Kaluza, \emph{On the unification problem in physics},
  \href{https://doi.org/10.1142/s0218271818700017}{\emph{International Journal
  of Modern Physics D} {\bfseries 27} (2018) 1870001}.

\bibitem{Klein:1926tv}
O.~Klein, \emph{{Quantum Theory and Five-Dimensional Theory of Relativity. (In
  German and English)}}, \href{https://doi.org/10.1007/BF01397481}{\emph{Z.
  Phys.} {\bfseries 37} (1926) 895}.

\bibitem{Cremmer:1978ds}
E.~Cremmer and B.~Julia, \emph{{The N=8 Supergravity Theory. 1. The
  Lagrangian}}, \href{https://doi.org/10.1016/0370-2693(78)90303-9}{\emph{Phys.
  Lett. B} {\bfseries 80} (1978) 48}.

\bibitem{Gibbons:1987ps}
G.W.~Gibbons and K.-i.~Maeda, \emph{{Black Holes and Membranes in Higher
  Dimensional Theories with Dilaton Fields}},
  \href{https://doi.org/10.1016/0550-3213(88)90006-5}{\emph{Nucl. Phys. B}
  {\bfseries 298} (1988) 741}.

\bibitem{Garfinkle:1990qj}
D.~Garfinkle, G.T.~Horowitz and A.~Strominger, \emph{{Charged black holes in
  string theory}}, \href{https://doi.org/10.1103/PhysRevD.43.3140}{\emph{Phys.
  Rev. D} {\bfseries 43} (1991) 3140}.

\bibitem{Delgado:2016zxv}
J.F.M.~Delgado, C.A.R.~Herdeiro and E.~Radu, \emph{{Violations of the Kerr and
  Reissner-Nordstr\"om bounds: Horizon versus asymptotic quantities}},
  \href{https://doi.org/10.1103/PhysRevD.94.024006}{\emph{Phys. Rev. D}
  {\bfseries 94} (2016) 024006}
  [\href{https://arxiv.org/abs/1606.07900}{{\ttfamily 1606.07900}}].

\bibitem{Ferrari:2000ep}
V.~Ferrari, M.~Pauri and F.~Piazza, \emph{{Quasinormal modes of charged,
  dilaton black holes}},
  \href{https://doi.org/10.1103/PhysRevD.63.064009}{\emph{Phys. Rev. D}
  {\bfseries 63} (2001) 064009}
  [\href{https://arxiv.org/abs/gr-qc/0005125}{{\ttfamily gr-qc/0005125}}].

\bibitem{Zhang:2015jda}
C.-Y.~Zhang, S.-J.~Zhang and B.~Wang, \emph{{Charged scalar perturbations
  around Garfinkle\textendash{}Horowitz\textendash{}Strominger black holes}},
  \href{https://doi.org/10.1016/j.nuclphysb.2015.07.030}{\emph{Nucl. Phys. B}
  {\bfseries 899} (2015) 37}
  [\href{https://arxiv.org/abs/1501.03260}{{\ttfamily 1501.03260}}].

\bibitem{Brito:2018hjh}
R.~Brito and C.~Pacilio, \emph{{Quasinormal modes of weakly charged
  Einstein-Maxwell-dilaton black holes}},
  \href{https://doi.org/10.1103/PhysRevD.98.104042}{\emph{Phys. Rev. D}
  {\bfseries 98} (2018) 104042}
  [\href{https://arxiv.org/abs/1807.09081}{{\ttfamily 1807.09081}}].

\bibitem{Blazquez-Salcedo:2019nwd}
J.L.~Bl\'azquez-Salcedo, S.~Kahlen and J.~Kunz, \emph{{Quasinormal modes of
  dilatonic Reissner\textendash{}Nordstr\"om black holes}},
  \href{https://doi.org/10.1140/epjc/s10052-019-7535-4}{\emph{Eur. Phys. J. C}
  {\bfseries 79} (2019) 1021}
  [\href{https://arxiv.org/abs/1911.01943}{{\ttfamily 1911.01943}}].

\bibitem{Astefanesei:2019qsg}
D.~Astefanesei, J.L.~Bl\'azquez-Salcedo, C.~Herdeiro, E.~Radu and
  N.~Sanchis-Gual, \emph{{Dynamically and thermodynamically stable black holes
  in Einstein-Maxwell-dilaton gravity}},
  \href{https://doi.org/10.1007/JHEP07(2020)063}{\emph{JHEP} {\bfseries 07}
  (2020) 063} [\href{https://arxiv.org/abs/1912.02192}{{\ttfamily
  1912.02192}}].

\bibitem{Zhang:2021edm}
C.-Y.~Zhang, P.~Liu, Y.~Liu, C.~Niu and B.~Wang, \emph{{Evolution of
  anti\textendash{}de Sitter black holes in Einstein-Maxwell-dilaton theory}},
  \href{https://doi.org/10.1103/PhysRevD.105.024010}{\emph{Phys. Rev. D}
  {\bfseries 105} (2022) 024010}
  [\href{https://arxiv.org/abs/2104.07281}{{\ttfamily 2104.07281}}].

\bibitem{Zhang:2021ybj}
C.-Y.~Zhang, P.~Liu, Y.~Liu, C.~Niu and B.~Wang, \emph{{Dynamical scalarization
  in Einstein-Maxwell-dilaton theory}},
  \href{https://doi.org/10.1103/PhysRevD.105.024073}{\emph{Phys. Rev. D}
  {\bfseries 105} (2022) 024073}
  [\href{https://arxiv.org/abs/2111.10744}{{\ttfamily 2111.10744}}].

\bibitem{Hirschmann:2017psw}
E.W.~Hirschmann, L.~Lehner, S.L.~Liebling and C.~Palenzuela, \emph{{Black Hole
  Dynamics in Einstein-Maxwell-Dilaton Theory}},
  \href{https://doi.org/10.1103/PhysRevD.97.064032}{\emph{Phys. Rev. D}
  {\bfseries 97} (2018) 064032}
  [\href{https://arxiv.org/abs/1706.09875}{{\ttfamily 1706.09875}}].

\bibitem{Julie:2017rpw}
F.-L.~Juli\'e, \emph{{On the motion of hairy black holes in
  Einstein-Maxwell-dilaton theories}},
  \href{https://doi.org/10.1088/1475-7516/2018/01/026}{\emph{JCAP} {\bfseries
  01} (2018) 026} [\href{https://arxiv.org/abs/1711.10769}{{\ttfamily
  1711.10769}}].

\bibitem{Khalil:2018aaj}
M.~Khalil, N.~Sennett, J.~Steinhoff, J.~Vines and A.~Buonanno, \emph{{Hairy
  binary black holes in Einstein-Maxwell-dilaton theory and their
  effective-one-body description}},
  \href{https://doi.org/10.1103/PhysRevD.98.104010}{\emph{Phys. Rev. D}
  {\bfseries 98} (2018) 104010}
  [\href{https://arxiv.org/abs/1809.03109}{{\ttfamily 1809.03109}}].

\bibitem{DeWolfe:2010he}
O.~DeWolfe, S.S.~Gubser and C.~Rosen, \emph{{A holographic critical point}},
  \href{https://doi.org/10.1103/PhysRevD.83.086005}{\emph{Phys. Rev. D}
  {\bfseries 83} (2011) 086005}
  [\href{https://arxiv.org/abs/1012.1864}{{\ttfamily 1012.1864}}].

\bibitem{DeWolfe:2011ts}
O.~DeWolfe, S.S.~Gubser and C.~Rosen, \emph{{Dynamic critical phenomena at a
  holographic critical point}},
  \href{https://doi.org/10.1103/PhysRevD.84.126014}{\emph{Phys. Rev. D}
  {\bfseries 84} (2011) 126014}
  [\href{https://arxiv.org/abs/1108.2029}{{\ttfamily 1108.2029}}].

\bibitem{Cai:2012xh}
R.-G.~Cai, S.~He and D.~Li, \emph{{A hQCD model and its phase diagram in
  Einstein-Maxwell-Dilaton system}},
  \href{https://doi.org/10.1007/JHEP03(2012)033}{\emph{JHEP} {\bfseries 03}
  (2012) 033} [\href{https://arxiv.org/abs/1201.0820}{{\ttfamily 1201.0820}}].

\bibitem{He:2013qq}
S.~He, S.-Y.~Wu, Y.~Yang and P.-H.~Yuan, \emph{{Phase Structure in a Dynamical
  Soft-Wall Holographic QCD Model}},
  \href{https://doi.org/10.1007/JHEP04(2013)093}{\emph{JHEP} {\bfseries 04}
  (2013) 093} [\href{https://arxiv.org/abs/1301.0385}{{\ttfamily 1301.0385}}].

\bibitem{Rougemont:2015ona}
R.~Rougemont, J.~Noronha and J.~Noronha-Hostler, \emph{{Suppression of baryon
  diffusion and transport in a baryon rich strongly coupled quark-gluon
  plasma}}, \href{https://doi.org/10.1103/PhysRevLett.115.202301}{\emph{Phys.
  Rev. Lett.} {\bfseries 115} (2015) 202301}
  [\href{https://arxiv.org/abs/1507.06972}{{\ttfamily 1507.06972}}].

\bibitem{Knaute:2017lll}
J.~Knaute and B.~K\"ampfer, \emph{{Holographic Entanglement Entropy in the QCD
  Phase Diagram with a Critical Point}},
  \href{https://doi.org/10.1103/PhysRevD.96.106003}{\emph{Phys. Rev. D}
  {\bfseries 96} (2017) 106003}
  [\href{https://arxiv.org/abs/1706.02647}{{\ttfamily 1706.02647}}].

\bibitem{Giataganas:2017koz}
D.~Giataganas, U.~G\"ursoy and J.F.~Pedraza, \emph{{Strongly-coupled
  anisotropic gauge theories and holography}},
  \href{https://doi.org/10.1103/PhysRevLett.121.121601}{\emph{Phys. Rev. Lett.}
  {\bfseries 121} (2018) 121601}
  [\href{https://arxiv.org/abs/1708.05691}{{\ttfamily 1708.05691}}].

\bibitem{Cai:2022omk}
R.-G.~Cai, S.~He, L.~Li and Y.-X.~Wang, \emph{{Probing QCD critical point and
  induced gravitational wave by black hole physics}},
  \href{https://doi.org/10.1103/PhysRevD.106.L121902}{\emph{Phys. Rev. D}
  {\bfseries 106} (2022) L121902}
  [\href{https://arxiv.org/abs/2201.02004}{{\ttfamily 2201.02004}}].

\bibitem{Myung:2018vug}
Y.S.~Myung and D.-C.~Zou, \emph{{Instability of
  Reissner\textendash{}Nordstr\"om black hole in Einstein-Maxwell-scalar
  theory}}, \href{https://doi.org/10.1140/epjc/s10052-019-6792-6}{\emph{Eur.
  Phys. J. C} {\bfseries 79} (2019) 273}
  [\href{https://arxiv.org/abs/1808.02609}{{\ttfamily 1808.02609}}].

\bibitem{Myung:2018jvi}
Y.S.~Myung and D.-C.~Zou, \emph{{Quasinormal modes of scalarized black holes in
  the Einstein\textendash{}Maxwell\textendash{}Scalar theory}},
  \href{https://doi.org/10.1016/j.physletb.2019.01.046}{\emph{Phys. Lett. B}
  {\bfseries 790} (2019) 400}
  [\href{https://arxiv.org/abs/1812.03604}{{\ttfamily 1812.03604}}].

\bibitem{Guo:2021zed}
G.~Guo, P.~Wang, H.~Wu and H.~Yang, \emph{{Scalarized
  Einstein\textendash{}Maxwell-scalar black holes in anti-de Sitter
  spacetime}},
  \href{https://doi.org/10.1140/epjc/s10052-021-09614-7}{\emph{Eur. Phys. J. C}
  {\bfseries 81} (2021) 864}
  [\href{https://arxiv.org/abs/2102.04015}{{\ttfamily 2102.04015}}].

\bibitem{Herdeiro:2018wub}
C.A.R.~Herdeiro, E.~Radu, N.~Sanchis-Gual and J.A.~Font, \emph{{Spontaneous
  Scalarization of Charged Black Holes}},
  \href{https://doi.org/10.1103/PhysRevLett.121.101102}{\emph{Phys. Rev. Lett.}
  {\bfseries 121} (2018) 101102}
  [\href{https://arxiv.org/abs/1806.05190}{{\ttfamily 1806.05190}}].

\bibitem{Fernandes:2019rez}
P.G.S.~Fernandes, C.A.R.~Herdeiro, A.M.~Pombo, E.~Radu and N.~Sanchis-Gual,
  \emph{{Spontaneous Scalarisation of Charged Black Holes: Coupling Dependence
  and Dynamical Features}},
  \href{https://doi.org/10.1088/1361-6382/ab23a1}{\emph{Class. Quant. Grav.}
  {\bfseries 36} (2019) 134002}
  [\href{https://arxiv.org/abs/1902.05079}{{\ttfamily 1902.05079}}].

\bibitem{Fernandes:2019kmh}
P.G.S.~Fernandes, C.A.R.~Herdeiro, A.M.~Pombo, E.~Radu and N.~Sanchis-Gual,
  \emph{{Charged black holes with axionic-type couplings: Classes of solutions
  and dynamical scalarization}},
  \href{https://doi.org/10.1103/PhysRevD.100.084045}{\emph{Phys. Rev. D}
  {\bfseries 100} (2019) 084045}
  [\href{https://arxiv.org/abs/1908.00037}{{\ttfamily 1908.00037}}].

\bibitem{Zhang:2021etr}
C.-Y.~Zhang, P.~Liu, Y.~Liu, C.~Niu and B.~Wang, \emph{{Dynamical charged black
  hole spontaneous scalarization in anti\textendash{}de Sitter spacetimes}},
  \href{https://doi.org/10.1103/PhysRevD.104.084089}{\emph{Phys. Rev. D}
  {\bfseries 104} (2021) 084089}
  [\href{https://arxiv.org/abs/2103.13599}{{\ttfamily 2103.13599}}].

\bibitem{Luo:2022roz}
W.-K.~Luo, C.-Y.~Zhang, P.~Liu, C.~Niu and B.~Wang, \emph{{Dynamical
  spontaneous scalarization in Einstein-Maxwell-scalar models in
  anti\textendash{}de Sitter spacetime}},
  \href{https://doi.org/10.1103/PhysRevD.106.064036}{\emph{Phys. Rev. D}
  {\bfseries 106} (2022) 064036}
  [\href{https://arxiv.org/abs/2206.05690}{{\ttfamily 2206.05690}}].

\bibitem{Doneva:2017bvd}
D.D.~Doneva and S.S.~Yazadjiev, \emph{{New Gauss-Bonnet Black Holes with
  Curvature-Induced Scalarization in Extended Scalar-Tensor Theories}},
  \href{https://doi.org/10.1103/PhysRevLett.120.131103}{\emph{Phys. Rev. Lett.}
  {\bfseries 120} (2018) 131103}
  [\href{https://arxiv.org/abs/1711.01187}{{\ttfamily 1711.01187}}].

\bibitem{Silva:2017uqg}
H.O.~Silva, J.~Sakstein, L.~Gualtieri, T.P.~Sotiriou and E.~Berti,
  \emph{{Spontaneous scalarization of black holes and compact stars from a
  Gauss-Bonnet coupling}},
  \href{https://doi.org/10.1103/PhysRevLett.120.131104}{\emph{Phys. Rev. Lett.}
  {\bfseries 120} (2018) 131104}
  [\href{https://arxiv.org/abs/1711.02080}{{\ttfamily 1711.02080}}].

\bibitem{Antoniou:2017acq}
G.~Antoniou, A.~Bakopoulos and P.~Kanti, \emph{{Evasion of No-Hair Theorems and
  Novel Black-Hole Solutions in Gauss-Bonnet Theories}},
  \href{https://doi.org/10.1103/PhysRevLett.120.131102}{\emph{Phys. Rev. Lett.}
  {\bfseries 120} (2018) 131102}
  [\href{https://arxiv.org/abs/1711.03390}{{\ttfamily 1711.03390}}].

\bibitem{Cunha:2019dwb}
P.V.P.~Cunha, C.A.R.~Herdeiro and E.~Radu, \emph{{Spontaneously Scalarized Kerr
  Black Holes in Extended Scalar-Tensor\textendash{}Gauss-Bonnet Gravity}},
  \href{https://doi.org/10.1103/PhysRevLett.123.011101}{\emph{Phys. Rev. Lett.}
  {\bfseries 123} (2019) 011101}
  [\href{https://arxiv.org/abs/1904.09997}{{\ttfamily 1904.09997}}].

\bibitem{Dima:2020yac}
A.~Dima, E.~Barausse, N.~Franchini and T.P.~Sotiriou, \emph{{Spin-induced black
  hole spontaneous scalarization}},
  \href{https://doi.org/10.1103/PhysRevLett.125.231101}{\emph{Phys. Rev. Lett.}
  {\bfseries 125} (2020) 231101}
  [\href{https://arxiv.org/abs/2006.03095}{{\ttfamily 2006.03095}}].

\bibitem{Herdeiro:2020wei}
C.A.R.~Herdeiro, E.~Radu, H.O.~Silva, T.P.~Sotiriou and N.~Yunes,
  \emph{{Spin-induced scalarized black holes}},
  \href{https://doi.org/10.1103/PhysRevLett.126.011103}{\emph{Phys. Rev. Lett.}
  {\bfseries 126} (2021) 011103}
  [\href{https://arxiv.org/abs/2009.03904}{{\ttfamily 2009.03904}}].

\bibitem{Berti:2020kgk}
E.~Berti, L.G.~Collodel, B.~Kleihaus and J.~Kunz, \emph{{Spin-induced
  black-hole scalarization in Einstein-scalar-Gauss-Bonnet theory}},
  \href{https://doi.org/10.1103/PhysRevLett.126.011104}{\emph{Phys. Rev. Lett.}
  {\bfseries 126} (2021) 011104}
  [\href{https://arxiv.org/abs/2009.03905}{{\ttfamily 2009.03905}}].

\bibitem{Damour:1993hw}
T.~Damour and G.~Esposito-Farese, \emph{{Nonperturbative strong field effects
  in tensor - scalar theories of gravitation}},
  \href{https://doi.org/10.1103/PhysRevLett.70.2220}{\emph{Phys. Rev. Lett.}
  {\bfseries 70} (1993) 2220}.

\bibitem{Brihaye:2019puo}
Y.~Brihaye and B.~Hartmann, \emph{{Spontaneous scalarization of boson stars}},
  \href{https://doi.org/10.1007/JHEP09(2019)049}{\emph{JHEP} {\bfseries 09}
  (2019) 049} [\href{https://arxiv.org/abs/1903.10471}{{\ttfamily
  1903.10471}}].

\bibitem{Peng:2019qrl}
Y.~Peng, \emph{{Scalarization of compact stars in the scalar-Gauss-Bonnet
  gravity}}, \href{https://doi.org/10.1007/JHEP12(2019)064}{\emph{JHEP}
  {\bfseries 12} (2019) 064}
  [\href{https://arxiv.org/abs/1910.13718}{{\ttfamily 1910.13718}}].

\bibitem{Blazquez-Salcedo:2020nhs}
J.L.~Bl\'azquez-Salcedo, C.A.R.~Herdeiro, J.~Kunz, A.M.~Pombo and E.~Radu,
  \emph{{Einstein-Maxwell-scalar black holes: the hot, the cold and the bald}},
  \href{https://doi.org/10.1016/j.physletb.2020.135493}{\emph{Phys. Lett. B}
  {\bfseries 806} (2020) 135493}
  [\href{https://arxiv.org/abs/2002.00963}{{\ttfamily 2002.00963}}].

\bibitem{LuisBlazquez-Salcedo:2020rqp}
J.~Luis Bl\'azquez-Salcedo, C.A.R.~Herdeiro, S.~Kahlen, J.~Kunz, A.M.~Pombo and
  E.~Radu, \emph{{Quasinormal modes of hot, cold and bald
  Einstein\textendash{}Maxwell-scalar black holes}},
  \href{https://doi.org/10.1140/epjc/s10052-021-08952-w}{\emph{Eur. Phys. J. C}
  {\bfseries 81} (2021) 155}
  [\href{https://arxiv.org/abs/2008.11744}{{\ttfamily 2008.11744}}].

\bibitem{Zhang:2021nnn}
C.-Y.~Zhang, Q.~Chen, Y.~Liu, W.-K.~Luo, Y.~Tian and B.~Wang, \emph{{Critical
  Phenomena in Dynamical Scalarization of Charged Black Holes}},
  \href{https://doi.org/10.1103/PhysRevLett.128.161105}{\emph{Phys. Rev. Lett.}
  {\bfseries 128} (2022) 161105}
  [\href{https://arxiv.org/abs/2112.07455}{{\ttfamily 2112.07455}}].

\bibitem{Zhang:2022cmu}
C.-Y.~Zhang, Q.~Chen, Y.~Liu, W.-K.~Luo, Y.~Tian and B.~Wang, \emph{{Dynamical
  transitions in scalarization and descalarization through black hole
  accretion}}, \href{https://doi.org/10.1103/PhysRevD.106.L061501}{\emph{Phys.
  Rev. D} {\bfseries 106} (2022) L061501}
  [\href{https://arxiv.org/abs/2204.09260}{{\ttfamily 2204.09260}}].

\bibitem{Jiang:2023yyn}
J.-Y.~Jiang, Q.~Chen, Y.~Liu, Y.~Tian, W.~Xiong, C.-Y.~Zhang et~al.,
  \emph{{Type I critical dynamical scalarization and descalarization in
  Einstein-Maxwell-scalar theory}},
  \href{https://arxiv.org/abs/2306.10371}{{\ttfamily 2306.10371}}.

\bibitem{Chen:2023eru}
Q.~Chen, Z.~Ning, Y.~Tian, B.~Wang and C.-Y.~Zhang, \emph{{Nonlinear dynamics
  of hot, cold and bald Einstein-Maxwell-scalar black holes in AdS spacetime}},
   \href{https://arxiv.org/abs/2307.03060}{{\ttfamily 2307.03060}}.

\bibitem{Choptuik:1992jv}
M.W.~Choptuik, \emph{{Universality and scaling in gravitational collapse of a
  massless scalar field}},
  \href{https://doi.org/10.1103/PhysRevLett.70.9}{\emph{Phys. Rev. Lett.}
  {\bfseries 70} (1993) 9}.

\bibitem{Choptuik:1996yg}
M.W.~Choptuik, T.~Chmaj and P.~Bizon, \emph{{Critical behavior in gravitational
  collapse of a Yang-Mills field}},
  \href{https://doi.org/10.1103/PhysRevLett.77.424}{\emph{Phys. Rev. Lett.}
  {\bfseries 77} (1996) 424}
  [\href{https://arxiv.org/abs/gr-qc/9603051}{{\ttfamily gr-qc/9603051}}].

\bibitem{Brady:1997fj}
P.R.~Brady, C.M.~Chambers and S.M.C.V.~Goncalves, \emph{{Phases of massive
  scalar field collapse}},
  \href{https://doi.org/10.1103/PhysRevD.56.R6057}{\emph{Phys. Rev. D}
  {\bfseries 56} (1997) R6057}
  [\href{https://arxiv.org/abs/gr-qc/9709014}{{\ttfamily gr-qc/9709014}}].

\bibitem{Bizon:1998kq}
P.~Bizon and T.~Chmaj, \emph{{Critical collapse of Skyrmions}},
  \href{https://doi.org/10.1103/PhysRevD.58.041501}{\emph{Phys. Rev. D}
  {\bfseries 58} (1998) 041501}
  [\href{https://arxiv.org/abs/gr-qc/9801012}{{\ttfamily gr-qc/9801012}}].

\bibitem{Gundlach:2007gc}
C.~Gundlach and J.M.~Martin-Garcia, \emph{{Critical phenomena in gravitational
  collapse}}, \href{https://doi.org/10.12942/lrr-2007-5}{\emph{Living Rev.
  Rel.} {\bfseries 10} (2007) 5}
  [\href{https://arxiv.org/abs/0711.4620}{{\ttfamily 0711.4620}}].

\bibitem{Liu:2022fxy}
Y.~Liu, C.-Y.~Zhang, Q.~Chen, Z.~Cao, Y.~Tian and B.~Wang, \emph{{The critical
  scalarization and descalarization of black holes in a generalized
  scalar-tensor theory}},  \href{https://arxiv.org/abs/2208.07548}{{\ttfamily
  2208.07548}}.

\bibitem{Janik:2006ft}
R.A.~Janik, \emph{{Viscous plasma evolution from gravity using AdS/CFT}},
  \href{https://doi.org/10.1103/PhysRevLett.98.022302}{\emph{Phys. Rev. Lett.}
  {\bfseries 98} (2007) 022302}
  [\href{https://arxiv.org/abs/hep-th/0610144}{{\ttfamily hep-th/0610144}}].

\bibitem{Chesler:2008hg}
P.M.~Chesler and L.G.~Yaffe, \emph{{Horizon formation and far-from-equilibrium
  isotropization in supersymmetric Yang-Mills plasma}},
  \href{https://doi.org/10.1103/PhysRevLett.102.211601}{\emph{Phys. Rev. Lett.}
  {\bfseries 102} (2009) 211601}
  [\href{https://arxiv.org/abs/0812.2053}{{\ttfamily 0812.2053}}].

\bibitem{Bhattacharyya:2009uu}
S.~Bhattacharyya and S.~Minwalla, \emph{{Weak Field Black Hole Formation in
  Asymptotically AdS Spacetimes}},
  \href{https://doi.org/10.1088/1126-6708/2009/09/034}{\emph{JHEP} {\bfseries
  09} (2009) 034} [\href{https://arxiv.org/abs/0904.0464}{{\ttfamily
  0904.0464}}].

\bibitem{Chesler:2009cy}
P.M.~Chesler and L.G.~Yaffe, \emph{{Boost invariant flow, black hole formation,
  and far-from-equilibrium dynamics in N = 4 supersymmetric Yang-Mills
  theory}}, \href{https://doi.org/10.1103/PhysRevD.82.026006}{\emph{Phys. Rev.
  D} {\bfseries 82} (2010) 026006}
  [\href{https://arxiv.org/abs/0906.4426}{{\ttfamily 0906.4426}}].

\bibitem{Buchel:2014gta}
A.~Buchel, R.C.~Myers and A.~van Niekerk, \emph{{Nonlocal probes of
  thermalization in holographic quenches with spectral methods}},
  \href{https://doi.org/10.1007/JHEP02(2015)017}{\emph{JHEP} {\bfseries 02}
  (2015) 017} [\href{https://arxiv.org/abs/1410.6201}{{\ttfamily 1410.6201}}].

\bibitem{Bhaseen:2012gg}
M.J.~Bhaseen, J.P.~Gauntlett, B.D.~Simons, J.~Sonner and T.~Wiseman,
  \emph{{Holographic Superfluids and the Dynamics of Symmetry Breaking}},
  \href{https://doi.org/10.1103/PhysRevLett.110.015301}{\emph{Phys. Rev. Lett.}
  {\bfseries 110} (2013) 015301}
  [\href{https://arxiv.org/abs/1207.4194}{{\ttfamily 1207.4194}}].

\bibitem{Basu:2012gg}
P.~Basu, D.~Das, S.R.~Das and T.~Nishioka, \emph{{Quantum Quench Across a Zero
  Temperature Holographic Superfluid Transition}},
  \href{https://doi.org/10.1007/JHEP03(2013)146}{\emph{JHEP} {\bfseries 03}
  (2013) 146} [\href{https://arxiv.org/abs/1211.7076}{{\ttfamily 1211.7076}}].

\bibitem{Gao:2012aw}
X.~Gao, A.M.~Garcia-Garcia, H.B.~Zeng and H.-Q.~Zhang, \emph{{Normal modes and
  time evolution of a holographic superconductor after a quantum quench}},
  \href{https://doi.org/10.1007/JHEP06(2014)019}{\emph{JHEP} {\bfseries 06}
  (2014) 019} [\href{https://arxiv.org/abs/1212.1049}{{\ttfamily 1212.1049}}].

\bibitem{Garcia-Garcia:2013rha}
A.M.~Garc\'\i{}a-Garc\'\i{}a, H.B.~Zeng and H.Q.~Zhang, \emph{{A thermal quench
  induces spatial inhomogeneities in a holographic superconductor}},
  \href{https://doi.org/10.1007/JHEP07(2014)096}{\emph{JHEP} {\bfseries 07}
  (2014) 096} [\href{https://arxiv.org/abs/1308.5398}{{\ttfamily 1308.5398}}].

\bibitem{Bai:2014tla}
X.~Bai, B.-H.~Lee, L.~Li, J.-R.~Sun and H.-Q.~Zhang, \emph{{Time Evolution of
  Entanglement Entropy in Quenched Holographic Superconductors}},
  \href{https://doi.org/10.1007/JHEP04(2015)066}{\emph{JHEP} {\bfseries 04}
  (2015) 066} [\href{https://arxiv.org/abs/1412.5500}{{\ttfamily 1412.5500}}].

\bibitem{Rangamani:2015agy}
M.~Rangamani, M.~Rozali and A.~Vincart-Emard, \emph{{Dynamics of Holographic
  Entanglement Entropy Following a Local Quench}},
  \href{https://doi.org/10.1007/JHEP04(2016)069}{\emph{JHEP} {\bfseries 04}
  (2016) 069} [\href{https://arxiv.org/abs/1512.03478}{{\ttfamily
  1512.03478}}].

\bibitem{Chen:2022vag}
Q.~Chen, Z.~Ning, Y.~Tian, B.~Wang and C.-Y.~Zhang, \emph{{Descalarization by
  quenching charged hairy black hole in asymptotically AdS spacetime}},
  \href{https://doi.org/10.1007/JHEP01(2023)062}{\emph{JHEP} {\bfseries 01}
  (2023) 062} [\href{https://arxiv.org/abs/2210.14539}{{\ttfamily
  2210.14539}}].

\bibitem{Chen:2022tfy}
Q.~Chen, Y.~Liu, Y.~Tian, X.~Wu and H.~Zhang, \emph{{Quench Dynamics in
  Holographic First-Order Phase Transition}},
  \href{https://arxiv.org/abs/2211.11291}{{\ttfamily 2211.11291}}.

\bibitem{Li:2013fhw}
W.-J.~Li, Y.~Tian and H.-b.~Zhang, \emph{{Periodically Driven Holographic
  Superconductor}}, \href{https://doi.org/10.1007/JHEP07(2013)030}{\emph{JHEP}
  {\bfseries 07} (2013) 030} [\href{https://arxiv.org/abs/1305.1600}{{\ttfamily
  1305.1600}}].

\bibitem{Yang:2023dvk}
P.~Yang, M.~Baggioli, Z.~Cai, Y.~Tian and H.~Zhang, \emph{{Holographic
  dissipative space-time supersolids}},
  \href{https://arxiv.org/abs/2304.02534}{{\ttfamily 2304.02534}}.

\bibitem{Breitenlohner:1982jf}
P.~Breitenlohner and D.Z.~Freedman, \emph{{Stability in Gauged Extended
  Supergravity}},
  \href{https://doi.org/10.1016/0003-4916(82)90116-6}{\emph{Annals Phys.}
  {\bfseries 144} (1982) 249}.

\bibitem{Chesler:2013lia}
P.M.~Chesler and L.G.~Yaffe, \emph{{Numerical solution of gravitational
  dynamics in asymptotically anti-de Sitter spacetimes}},
  \href{https://doi.org/10.1007/JHEP07(2014)086}{\emph{JHEP} {\bfseries 07}
  (2014) 086} [\href{https://arxiv.org/abs/1309.1439}{{\ttfamily 1309.1439}}].

\bibitem{Klebanov:1999tb}
I.R.~Klebanov and E.~Witten, \emph{{AdS / CFT correspondence and symmetry
  breaking}}, \href{https://doi.org/10.1016/S0550-3213(99)00387-9}{\emph{Nucl.
  Phys. B} {\bfseries 556} (1999) 89}
  [\href{https://arxiv.org/abs/hep-th/9905104}{{\ttfamily hep-th/9905104}}].

\bibitem{Abbott:1981ff}
L.F.~Abbott and S.~Deser, \emph{{Stability of Gravity with a Cosmological
  Constant}}, \href{https://doi.org/10.1016/0550-3213(82)90049-9}{\emph{Nucl.
  Phys. B} {\bfseries 195} (1982) 76}.

\bibitem{Gibbons:1976ue}
G.W.~Gibbons and S.W.~Hawking, \emph{{Action Integrals and Partition Functions
  in Quantum Gravity}},
  \href{https://doi.org/10.1103/PhysRevD.15.2752}{\emph{Phys. Rev. D}
  {\bfseries 15} (1977) 2752}.

\bibitem{Bianchi:2001kw}
M.~Bianchi, D.Z.~Freedman and K.~Skenderis, \emph{{Holographic
  renormalization}},
  \href{https://doi.org/10.1016/S0550-3213(02)00179-7}{\emph{Nucl. Phys. B}
  {\bfseries 631} (2002) 159}
  [\href{https://arxiv.org/abs/hep-th/0112119}{{\ttfamily hep-th/0112119}}].

\bibitem{Elvang:2016tzz}
H.~Elvang and M.~Hadjiantonis, \emph{{A Practical Approach to the
  Hamilton-Jacobi Formulation of Holographic Renormalization}},
  \href{https://doi.org/10.1007/JHEP06(2016)046}{\emph{JHEP} {\bfseries 06}
  (2016) 046} [\href{https://arxiv.org/abs/1603.04485}{{\ttfamily
  1603.04485}}].

\bibitem{Ishii:2022lwc}
T.~Ishii, Y.~Kaku and K.~Murata, \emph{{Energy extraction from AdS black holes
  via superradiance}},
  \href{https://doi.org/10.1007/JHEP10(2022)024}{\emph{JHEP} {\bfseries 10}
  (2022) 024} [\href{https://arxiv.org/abs/2207.03123}{{\ttfamily
  2207.03123}}].

\bibitem{Aprile:2009ai}
F.~Aprile and J.G.~Russo, \emph{{Models of Holographic superconductivity}},
  \href{https://doi.org/10.1103/PhysRevD.81.026009}{\emph{Phys. Rev. D}
  {\bfseries 81} (2010) 026009}
  [\href{https://arxiv.org/abs/0912.0480}{{\ttfamily 0912.0480}}].

\bibitem{Liu:2010ka}
Y.~Liu and Y.-W.~Sun, \emph{{Holographic Superconductors from
  Einstein-Maxwell-Dilaton Gravity}},
  \href{https://doi.org/10.1007/JHEP07(2010)099}{\emph{JHEP} {\bfseries 07}
  (2010) 099} [\href{https://arxiv.org/abs/1006.2726}{{\ttfamily 1006.2726}}].

\bibitem{Goldstein:2010aw}
K.~Goldstein, N.~Iizuka, S.~Kachru, S.~Prakash, S.P.~Trivedi and A.~Westphal,
  \emph{{Holography of Dyonic Dilaton Black Branes}},
  \href{https://doi.org/10.1007/JHEP10(2010)027}{\emph{JHEP} {\bfseries 10}
  (2010) 027} [\href{https://arxiv.org/abs/1007.2490}{{\ttfamily 1007.2490}}].

\bibitem{Cadoni:2011kv}
M.~Cadoni and P.~Pani, \emph{{Holography of charged dilatonic black branes at
  finite temperature}},
  \href{https://doi.org/10.1007/JHEP04(2011)049}{\emph{JHEP} {\bfseries 04}
  (2011) 049} [\href{https://arxiv.org/abs/1102.3820}{{\ttfamily 1102.3820}}].

\bibitem{Gouteraux:2011qh}
B.~Gouteraux, J.~Smolic, M.~Smolic, K.~Skenderis and M.~Taylor,
  \emph{{Holography for Einstein-Maxwell-dilaton theories from generalized
  dimensional reduction}},
  \href{https://doi.org/10.1007/JHEP01(2012)089}{\emph{JHEP} {\bfseries 01}
  (2012) 089} [\href{https://arxiv.org/abs/1110.2320}{{\ttfamily 1110.2320}}].

\bibitem{Salvio:2012at}
A.~Salvio, \emph{{Holographic Superfluids and Superconductors in
  Dilaton-Gravity}}, \href{https://doi.org/10.1007/JHEP09(2012)134}{\emph{JHEP}
  {\bfseries 09} (2012) 134} [\href{https://arxiv.org/abs/1207.3800}{{\ttfamily
  1207.3800}}].

\end{thebibliography}\endgroup

\end{document}